\documentclass[onecolumn]{emulateapj}

\usepackage{graphicx}
\usepackage{color}
\usepackage{epsfig}
\def\lsim{\raise0.3ex\hbox{$<$}\kern-0.75em{\lower0.65ex\hbox{$\sim$}}}
\def\gsim{\raise0.3ex\hbox{$>$}\kern-0.75em{\lower0.65ex\hbox{$\sim$}}}

\shorttitle{Extremely dense cores in Oph-A}
\shortauthors{Kawabe et al.}

\begin{document}

\title{Extremely Dense Cores associated with Chandra Sources in Ophiuchus A: Forming Brown Dwarfs Unveiled? \\
 }

\author{Ryohei Kawabe}
\affiliation{National Astronomical Observatory, Mitaka, Tokyo 181-8588, Japan}
\affiliation{The Graduate University for Advanced Studies (SOKENDAI), 2-21-1 Osawa, Mitaka, Tokyo 181-0015, Japan}
\affiliation{Department of Astronomy, School of Science, University of Tokyo, Bunkyo, Tokyo, 113-0033, Japan.}

\author{Chihomi Hara}
\affiliation{Department of Astronomy, Graduate School of Science, The University of Tokyo, Tokyo 113-0033, Japan}
\affiliation{NEC Corporation Radio Application, Guidance and Electro-Optics Division, 
1-10, Nisshin-cho, Fuchu, Tokyo 183-8501, Japan}

\author{Fumitaka Nakamura}
\affiliation{National Astronomical Observatory, Mitaka, Tokyo 181-8588, Japan}
\affiliation{The Graduate University for Advanced Studies (SOKENDAI), 2-21-1 Osawa, Mitaka, Tokyo 181-0015, Japan}
\affiliation{Department of Astronomy, School of Science, University of Tokyo, Bunkyo, Tokyo, 113-0033, Japan.}

\author{Kazuya Saigo}
\affiliation{National Astronomical Observatory, Mitaka, Tokyo 181-8588, Japan}

\author{Takeshi Kamazaki}
\affiliation{National Astronomical Observatory, Mitaka, Tokyo 181-8588, Japan}

\author{Yoshito Shimajiri}
\affiliation{Laboratoire AIM, CEA/DSM-CNRS-Universit$\acute{\rm e}$ Paris Diderot, IRFU/Service d’Astrophysique, CEA, Saclay, 91191 Gif-sur-Yvette, France.}

\author{KengoTomida}
\affiliation{Department of Earth and Space Science, Osaka University, Toyonaka, Osaka 560-0043}

\author{Shigehisa Takakuwa}
\affiliation{Department of Physics and Astronomy, Graduate School of Science and Engineering, Kagoshima University, 1-21-35 Korimoto, Kagoshima,
Kagoshima 890-0065, Japan}
\affiliation{Academia Sinica Institute of Astronomy and Astrophysics, P.O. Box 23-141, Tapei 10617, Taiwan.}

\author{Yohko Tsuboi}
\affiliation{Department of Physics, Faculty of Science and Engineering, Chuo University, 1-13-27 Kasuga, Bunkyo-ku, Tokyo 112-8551, Japan}

\author{Masahiro N. Machida}
\affiliation{Department of Earth and Planetary Science, Faculty of Science, Kyushu University, Hakozaki 6-10-1, Higashi-ku, Fukuoka, 812-8581, Japan.}

\author{James Di Francesco}
\affiliation{NRC Herzberg Inst of Astrophysics, 5071 W Saanich Rd, Victoria BC V9E 2E7, British Columbia, Canada } 

\author{Rachel Friesen}
\affiliation{Dunlap Institute for Astronomy and Astrophysics, the University of Toronto, 50 St George St, Toronto ON M5S 3H4 Canada } 
\affiliation{North American ALMA Science Center, National Radio Astronomy Observatory, 520 Edgemont Rd, Charlottesville VA 22903}

\author{Naomi Hirano}
\affiliation{Academia Sinica Institute of Astronomy and Astrophysics, P.O. Box 23-141, Tapei 10617, Taiwan.}

\author{Yumiko Oasa}
\affiliation{Faculty of Education, Saitama University, 255 Shimo-Okubo, Sakura-ku, Saitama 338-8570, Japan.} 

\author{Motohide Tamura}
\affiliation{Department of Astronomy, Graduate School of Science, The University of Tokyo, Tokyo 113-0033, Japan}

\author{Yoichi Tamura}
\affiliation{Division of Particle and Astrophysical Science, Graduate School of Science, Nagoya University, 
Aichi 464-8602, Japan.}

\author{Takashi Tsukagoshi}
\affiliation{College of Science, Ibaraki University, Bunkyo 2-1-1, Mito, Ibaraki 310-8512, Japan.}

\author{David Wilner}
\affiliation{Harvard-Smithsonian Center for Astrophysics, 60 Garden Street, Cambridge, MA 02
138, USA.}

\begin{abstract}
On the basis of various data such as ALMA, JVLA, Chandra, {\it Herschel}, and {\it Spitzer}, we confirmed that two protostellar candidates in Oph-A
are bona fide protostars or proto-brown dwarfs (proto-BDs) in extremely early evolutionary stages.
Both objects are barely visible across infrared  (IR, i.e., near-IR to far-IR) bands.  
The physical nature of the  cores is very similar to that expected 
in first hydrostatic cores (FHSCs), objects  theoretically predicted in the evolutionary phase prior to stellar core formation with gas densities of  
$\sim$ 10$^{11-12}$ cm$^{-3}$. This suggests that the evolutionary stage is close to the FHSC formation phase. 
The two objects are associated with faint X-ray sources, suggesting that they are in very early phase of stellar core formation with magnetic activity.
In addition, we  found the CO outflow components around both sources which may originate from the young outflows 
driven by these sources.
The masses of these objects are calculated to be $\sim 0.01-0.03$ $M_\odot$ from the dust continuum emission.
Their physical properties are consistent with that expected from the numerical model of forming brown dwarfs. 
These facts (the  X-ray detection, CO outflow association, and FHSC-like spectral energy distributions) strongly indicate that the two objects are proto-BDs 
or will be in the very early phase of protostars which will evolve more massive protostars if they gain enough mass from the surroundings.
The ages of these two objects are likely to be within $\sim 10^3$ years after the protostellar core (or second core) formation, 
taking into account the outflow dynamical times ($\lesssim$ 500 yrs).
\end{abstract}


\section{Introduction}
How a star forms is a long-standing basic question in astronomy that has been investigated 
extensively during the past five decades \citep{hayashi66,lar69,shu87,mckee07,andre09}. 
The widely-accepted standard scenario of the formation of stars ($M_* > $ 0.08 $M_\odot$) 
 is the gravitational collapse of a 
molecular cloud core, or so-called ``prestellar core''. Prior to the protostellar phase, 
the central densest part of the collapsing cloud becomes adiabatic as its density reaches 
$\sim 10^{-13}$ g cm$^{-3}$ and its temperature becomes high enough to pause the collapse \citep{lar69}. 
At this point, with the collapsed object in hydrostatic equilibrium, 
it is referred to as a {\it first hydrostatic core} or {\it first core} in short \citep{lar69,masu98}. 
Its mass and radius are, respectively, $\sim 0.05 \, M_\odot$ 
and $\sim$ 4 au with no rotation \citep{masu98} and up to $\sim$100 au 
with rotation \citep{bate02,matsu03,saigo08}.
Its predicted luminosity, 0.01$-$0.1 $L_\odot$ \citep{lar69,bate02},
is defined by its radius and the mass 
accretion rate onto its surface \citep{bate02}.
When the core gains enough mass by accreting the surrounding gas, it evolves into a protostar.
This standard scenario is sometimes called ``core accretion" model.
The core accretion model assumes that a protostar is located at the center of a dense core from which the protostar
obtains its mass, determining the final mass of a star formed.  
Another well-discussed scenario of the formation of stars is the competitive accretion model \citep{bonnell00}.
In this model, an initially very low-mass compact core called a stellar seed is created from the parent molecular clouds
and it gains additional mass from the surroundings through Bondi-Hoyle accretion.

On the other hand, substellar-mass objects such as brown dwarfs (BDs) and planetary-mass objects (PMOs)
(i.e., $<$ 0.08 $M_\odot$)  have no widely accepted scenario for their formation \citep{bate02,basu12,mckee07}. 
Two possibilities are presently under debate, 
i.e., the same formation pathway as stars or dynamical ejection in the star-forming disk. 
In the latter scenario, substellar objects can be formed in disks via gravitational fragmentation 
if the disks are massive and unstable; the substellar objects are then 
ejected through many-body interaction. 
Ejected objects do not accrete much material  because the timescale of fragmentation and 
ejection is short ($\sim$ 10$^{3-4}$ yrs), hence less massive objects are born. 
Ejected dense clumps could also form BDs or PMOs if they can survive tidal disruption 
by being extremely dense like first cores \citep{basu12, andre09}. 
Concrete examples of forming BDs are needed  for discriminating between these two scenarios.

However, the formation processes of protostars and substellar-mass objects in the earliest stage remains uncertain
because there are not enough observational examples. 
Thus, it is important to observationally identify protostars and proto-BDs that are in the very early evolutionary stages.

The $\rho$ Ophiuchi star-forming region is one of the best places for the study of formation of protostars and substellar-mass objects 
because of its proxmity.
Recent VLBA observations have provided  an accurate distance to the $\rho$ Oph star-forming region 
 (Lynds 1688) of  137 $\pm$ 1.2 pc \citep{ortiz17}. Previous distance estimates to $\rho$ Oph 
fall into the range of 120 $-$ 140 pc \citep{kunde98,luh09,loinard08,lombardi08}.
Hereafter, we adopt 137 pc.
Given $\rho$ Oph's close proximity, we can easily achieve spatial resolution comparable to
the size of our solar system ($\sim$ 100 au)  using currently available state-of-the-art facilities.

We observed the densest part of the Ophiuchus molecular cloud,
the $\rho$-Ophiuchi A region \citep{difrancesco04}, using ALMA  in 2015 
and JVLA in 2012. 
Oph A was first observed by \citet{ward-thompson89} in submillimeter continuum with UKIRT,
and they named the brightest submillimeter core  SM1. 
Later, \citet{andre93} revealed that Oph A contains several dust condensations along the filamentary ridge 
with JCMT and IRAM 30-m telescopes.
On the basis of the Nobeyama Millimeter Array (NMA) observations at 2 mm and 3 mm, 
\citet{kama01} found that SM1 includes 1$''$ ($\sim$100 au) scale 
bright condensations.  They named the brightest condensation Source A.
Source A is also identified in 850 $\mu$m and 1.3-mm by \citet{frie14} and \citet{naka12}, respectively.
It is also identified in N$_2$H$^+$ ($J=1-0$) as N3 by \citet{difrancesco04}. 
Hereafter, we refer to source-A as SM1-A, taking into account the previous identifications based on the dust emission.

In the present paper, we discuss the physical properties of two protostellar candidates 
in Oph-A 
from the ALMA and JVLA observations toward the region and also 
X-ray, IR and radio data of the $\rho$ Ohpiuchi star-forming region. 
The two are SM1-A and a dust continuum source (hereafter, referred to as
Source-X) located between SM1-A and VLA 1623 which is an archetypal Class-0 protostar \citep{andre93}. 
The former was identified with a protostellar candidate with faint X-ray emission  \citep{frie14}. 
The latter source was recently detected in 3-mm with ALMA by \citet{kirk17} as core No. 10.
Its compactness indicates that it is a protostellar candidate.  
It is worth noting that Source-X is visible as local peaks in the 350 $\mu$m and 450 $\mu$m maps 
of \citet{andre93}.
Here we propose that these two protostellar candidates are bona fide protostars or proto-brown dwarfs.

\section{Observations and Data Analysis}

We used the (sub)millimeter, infrared and X-ray observation data to identify the substellar-mass objects. 
The millimeter and submillimeter
data were mainly obtained by our ALMA Band-6 mosaic observations
in Cycle-2 and JVLA 41 GHz observations. We also analyzed ALMA archive
data, which were Cycle-0 Band-7 and Cycle-2 Band-6 data, for high
angular resolution imaging and determining spectral energy distribution (SED) of the
candidates. At the wavelengths of NIR, {\it Spitzer} IRAC and MIPS data were retrieved
to search for IR counterparts. 
Furthermore, we reanalyzed Chandra X-ray
data taken in 2000 and merged the new data taken in 2014 to improve the signal-to-noise ratios and attempt to
identify deeply embedded young stellar and substellar objects that cannot be seen even at infrared wavelengths.
In the following, we describe details of these observations. 
Parameters of (sub)millimeter observations are summarized in Table \ref{tab:interferometer}.

\subsection{SMA Observations} \label{subsec:sma}

SM1-A was observed with the SubMillimeter Array (SMA)\footnote{The SMA is a joint
project between the Smithsonian Astrophysical Observatory
and the Academia Sinica Institute of Astronomy and Astrophysics and is funded by the
Smithsonian Institution and the Academia Sinica.} on
2007 July 29 in its compact-north configuration over the hour angle coverage of
-1.4$^{h}$ to 4.2$^{h}$. Details of the SMA are described by \citet{ho04}.
These SMA data were originally taken for polarimetric measurements
(S.P. Lai, private communication). The continuum data were first published
in our previous paper \citep{naka12}, which describes
the details of the observing parameters.
Seven out of the eight SMA antennas were used, providing projected baseline
lengths from 6.8 m to 125.5 m. The atmospheric transparency was good,
with the 225 GHz opacity ranging from $\sim$0.05 to 0.09
measured at the nearby Caltech Submillimeter Observatory (CSO).
The double sideband (DSB) receivers were tuned with a local oscillator (LO) frequency of 340.8 GHz.
The IF frequency is 5 GHz, and in each sideband the correlator covers the 2 GHz bandwidth.
Observations of NRAO 530 were interleaved with the target for gain calibration,
whose absolute flux density at 340 GHz was measured to be 1.4 Jy by bootstrapping
from observations of Uranus. The absolute flux accuracy is $\sim$15 $\%$.
Strong quasars 3C273 and 3C454.3 were adopted as the passband calibrators.
The raw visibility data were calibrated with an IDL-based reduction package, MIR \citep{sco93},
and the calibrated visibility data were Fourier-transformed and CLEANed with MIRIAD \citep{sau95}.

After the normal calibration and imaging processes, a very bright compact source
was identified, which enabled us to perform self-calibration.
The phase-only self-calibration improved the signal-to-noise ratio of the continuum image
by $\sim$30$\%$ and sharpened the image.

\subsection{JVLA 41 GHz Observations} \label{subsec:jvla}

SM1-A was also observed at 41 GHz ($\lambda$ = 7.3 mm) with
the Karl G. Jansky VLA (JVLA), which consists of the 27 25-m antennas.
The observations were conducted
on 2012 September 3 in the B configuration, covering projected baseline
lengths from 80 m to 7.4 km.
The correlator was configured to have 16 spectral windows with 128 MHz bandwidth each,
providing in total $\sim$2 GHz bandwidth from 39.998-0.064 GHz to 41.884+0.064 GHz.
J1256-0547, J1625-2527, and 3C286 were used for bandpass, complex gain, and flux scale calibrations,
respectively. In the observing sequence, the fast switching mode between the target
and the gain calibrator J1625-2527 separated at $\sim$1$^\circ$ was adopted.
The total on-source integration time was 2880 seconds. The 7.3 mm continuum image was made
with natural weighting to maximize the signal-to-noise ratio, yielding a 
synthesized beam size of 0$\farcs$32 $\times$ 0$\farcs$15 (P.A.=11.4$\degr$).
The FOV of the JVLA 25-m antennas at 41 GHz is $\sim$73$\arcsec$, and hence
VLA 1623, separated by $\sim$35$\arcsec$ from SM1 SM1-A, was
observed simultaneously. VLA 1623 and knot-a (or VLA 1623B) were clearly detected
in our observations together with SM1-A, and knot-b (or VLA 1623W) was also detected
above a 3 $\sigma$ level.

\subsection{ALMA Observations and Data Reduction} \label{subsec:alma}

\subsubsection{Band-7 Observations in Cycle-0}

The 810 $\mu$m and 835 $\mu$m (372.4 GHz and 359.2 GHz, respectively) continuum and 
N$_2$H$^+$ $J=4-3$ (372 GHz) observations toward SM1-A 
were performed on 2012 August 24$-$25 in ALMA Cycle-0 \citep[see][]{frie14}. 
The array was in the Cycle-0 Extended configuration with projected baseline 
lengths between $\sim$ 26 m and $\sim$ 500 m and sensitive to maximum angular scales 
of $\sim$ $3''$. The field of view (FOV) was $\sim$ $18''$ in diameter. 
Using the 26 12-m antennas, the total on-source integration time was 2400 seconds 
 (4 hours including the overheads). The weather conditions were good during 
the observations, and the system noise temperature was in the range of 400 K to 550 K 
at 372 GHz, and $\sim$ 200 K at 359 GHz. 
The 810 $\mu$m and 835 $\mu$m continuum data were taken from the upper- and the lower sidebands 
of the Band-7 receivers, respectively,  with four sets of 128 MHz spectral windows of the ALMA 
baseline correlator. The line-free channels were summed to form continuum visibility 
data of $\sim$ 256 MHz in bandwidth each from the 810 $\mu$m and 835 $\mu$m data. 
J1517-243, J1625-254, and Titan were used for bandpass, complex gain, and flux scale calibrations, respectively.

Data reduction was conducted on the Common Astronomy 
Software Applications package (CASA) version 4.2.2. 
 In addition to the standard calibration procedure for ALMA data, 
 we tried self-calibration for each of the 810 $\mu$m and 835 $\mu$m continuum datasets, 
 since SM1-A, located near the center of the FOV, is bright enough 
 for such calibration to work well. 
 Only phase self-calibration was done for the data set. 
 For the 835 $\mu$m image, the self-calibration was quite successful, 
 and a signal-to-noise ratio of more than 600 was achieved. 
In the original image by \citet{frie14} who applied the uniform weighting to the $uv$ data to create the image,
the achieved rms noise level was 2.0 mJy beam$^{-1}$ with the FWHM beam size of $0\arcsec.45 \times 0\arcsec.34$.
By applying the self-calibaration with natural weighting, we could achieve the rms noise level of 
0.39 mJy beam$^{-1}$, which is very close to the theoretical thermal noise expectation.

\subsubsection{Band-6 Observations in Cycle-2}

We obtained 219 GHz continuum data from the ALMA archive.
The 219 GHz continuum observations toward VLA 1623 were obtained in 2015 
for ALMA Cycle-2 program (project code: 2013.1.01004.S; S.-P. Lai). 
 The continuum data were analyzed in the same way 
 as the above 810 $\mu$m and 835 $\mu$m data with self-calibration. 
 The obtained beam size is 0$\farcs$6 $\times 0\farcs 34$ and the rms noise achieved is 
 $\sim$ 0.084 mJy beam$^{-1}$ at the center of the FOV.   
Uncertainty of the absolute flux density scale is  $\sim$ 10 $\%$.

\subsubsection{Cycle-2 Mosaic Observations}

We obtained ALMA Cycle-2 observations at 226 GHz 
 using the 12-m Array  with 150 pointings and ACA with 58 pointings to cover 
2$\arcmin$ $\times$ 3$\arcmin$ region of Oph-A. 
We successfully obtained the combined images for the continuum and 
three isotopic CO ($J=2-1$) lines, i.e, $^{12}$CO($J=2-1$), 
$^{13}$CO ($J=2-1$), and C$^{18}$O ($J=2-1$). 
The details of the observed continuum and molecular lines are summarized in Table \ref{tab:line}. 
The observation parameters are the same as those of the Oph B2 observations 
\citep{kama18}.
The reference position of the 12-m Array and the 7-m Array
was set to ($\alpha _{J2000.0}$, $\delta _{J2000.0}$)
= (16$^{\mathrm h}$27$^{\mathrm m}$26\fs507, $-$24\arcdeg31\arcmin28\farcs63), which is
the same as that of \citet{kama18}.
The $uv$ ranges sampled in the 12-m Array and 7-m Array data were
12.5 $-$ 348 k$\lambda$ and 8.1 $-$ 48 k$\lambda$, respectively. The
minimum $uv$ distance of the combined data corresponds to 25\arcsec.
We used quasars, J1517-2422 and J1733-1304, for the bandpass calibrators adopted for the 12-m Array and 7-m Array 
observations, respectively. We observed a quasar J1625-2527 as the phase calibrator for both
arrays. We used Titan and Mars for the flux calibrations of the 12-m Array
and the 7-m Array observations, respectively. 
For the absolute  flux scale of the solar
system objects (i.e., flux calibrators), we used the Butler-JPL-Horizons 2012 model.

\begin{figure}
\epsscale{1.20}
\centering
\plotone{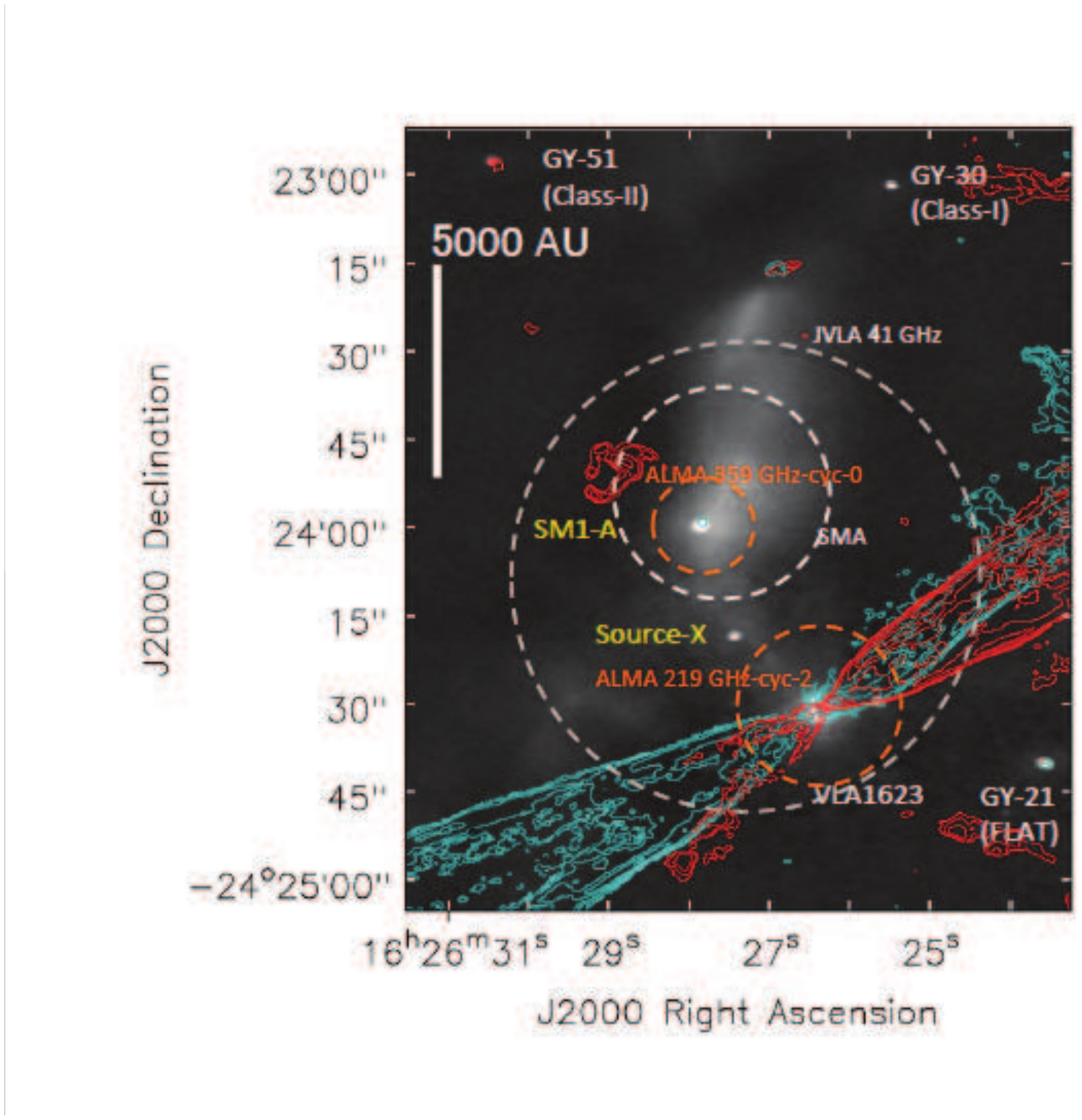}
\caption{Fields of view for JVLA, ALMA, and SMA single-pointing observations 
superposed on images of Oph A in 1.3 mm continuum (color) and {$^{12}$}CO ($J=2-1$) blue-shifted 
and red-shifted emission (contours) obtained from the ALMA mosaic observations 
with the ALMA 12-m and 7-m arrays.
The names of 1.3 mm detected YSOs are shown as well as two sources, SM1-A and Source-X. 
The detailed observational parameters are summarized in Table 1.}
\label{fig1}
\end{figure}

The 12-m Array and 7-m Array data were calibrated and imaged 
using the CASA pipeline version 4.2.2 and version 4.5.3, respectively. 
We modified the scripts prepared by the ALMA observatory in which the shadowing criterion for the 
bandpass data was reduced from 7 m to 6 m to recover some data flagged by the original scripts.
Then we conducted the calibration by ourselves.
In this paper, we will describe briefly a combined continuum map 
at 226 GHz and $^{12}$CO ($J=2-1$) detection toward two sources 
of interest for comparison purposes.   
The synthesized beams and the sensitivities of the dust continuum and $^{12}$CO data are 
summarized in Table \ref{tab:line}.
In Figure \ref{fig1}, we present the combined 12-m Array and 7-m Array 
1.3-mm continuum image of Oph A with
contours of blue-shifted and red-shifted $^{12}$CO ($J=2-1$) emission.
We also indicate the FOVs of the SMA, ALMA Cycle-0, ALMA 218 GHz Cycle-2, 
and JVLA observations with circles in Figure \ref{fig1}.

\subsection{Chandra Observations}

The Chandra X-ray observations were performed in May 2000 (Obs ID = 637) and in December 2014 (Obs ID =
17249), both of which include the Oph A region. The comprehensive
reports for the former dataset have been given by \citet{ima03} and \citet{gag04}. 
Both datasets have exposure of
$\sim$100 ks. 
The Oph A region was observed with the ACIS-I imaging array, which
is comprised of four charge-coupled devices (CCDs) that are front-side
illuminated (I0, I1, I2, and I3).

After downloading both of the datasets from the Chandra X-ray Center
(CXC), we did the data processing and analysis using the Chandra
Interactive Analysis of Observations (CIAO) software, developed by the
Chandra Science Center. We started the analysis of the latter dataset
(Obs ID = 17249) simply using the Level 2 events list provided by the
Chandra Science Center, which was filtered on the good time intervals
(GTIs), cosmic ray rejection, and position transformation to celestial
coordinates (RA, Dec) from the more primitive Level 1 outputs (see
more detail in the Chandra Analysis Guide
(http:cxc.harvard.edu/ciao/guides/). As for the former dataset (Obs ID
= 637), we generated a new Level 2 events list by reprocessing the
Level 1 output by ourselves, with updated calibration data.
We merged the data taken in 2000 and 2014 to improve the signal-to-noise
ratios, using the CIAO software, ``reproject\_obs".

\subsection{IR Data Analysis and Photometry} \label{subsec:ir}

\subsubsection{Spitzer IRAC/MIPS Data and Herschel Data} 

We retrieved 3.6 $\mu$m, 4.5 $\mu$m, 5.8 $\mu$m, 8.0 $\mu$m, 24 $\mu$m, and 70 $\mu$m 
images obtained with the Infrared Array 
Camera (IRAC; Fazio  et al. 2004) and the Multiband Imaging Photometer for {\it Spitzer} (MIPS) 
\citep {rie04} 
from the NASA/IPAC Infrared Science Archive
to search for the infrared counterparts of the submillimeter sources 
and measure the corresponding flux densities. 
The basic calibrated data (BCD) of IRAC and MIPS are processed through masking, 
flat fielding, background matching, and mosaicing using the MOPEX 
(MOsaicker and Point source EXtractor) software, which is a package developed 
at the Spitzer Science Center for astronomical image processing. 
We have also analyzed the recent {\it Herschel} 70 $\mu$m data from the Herschel Science Archive, 
which has a higher signal-to-noise ratio than the {\it Spitzer} 70 $\mu$m image.

\subsubsection{ Analysis of J, H, and Ks Band Data} 

We observed the 7$\arcmin$ $\times$ 7$\arcmin$ field toward SM1-A
in the near-infrared $J$, $H$, and $Ks$ bands simultaneously with the near-infrared camera 
SIRIUS on the InfraRed Survey Facility (IRSF) 1.4 m telescope at the Sutherland South African Astronomical 
Observatory on 2004 July 7, 9, 11, 16, and October 17 and 2005 March 15, April 30, 
May 17, and July 6. The total integration time was 7485 sec. 
We used NOAO's Imaging Reduction and Analysis Facility (IRAF) software package to carry out the data reduction.
We applied standard procedures of near-infrared array image reduction, including dark current subtraction, sky subtraction, and flat
fielding.
Photometric calibration was done with the 2MASS point sources in the same field. 
Identification and photometry of point sources in all frames were performed with using the DAOPHOT packages in IRAF.
The upper limit for source detection was $K_s > $19.5 mag (3 $\sigma$ upper limit).

\section{Results}

\subsection{Photometry}

We measured flux densities of SM1-A and Source-X using interferometric data available, 
{\it Spitzer}, and {\it Herschel} data, and have summarized them in Table \ref{tab:properties} together 
with number of counts by Chandra. The flux densities are measured in 
the $\sim$ 2$\arcsec$ $\times$ 2$\arcsec$ regions of spatially resolved cases 
to exclude extended emission which can be 
seen in the SMA visibility distribution (see also Section 3.4). 
Peak flux densities 
in units of mJy beam$^{-1}$ are given for unresolved cases. Upper limits 
of 3 $\sigma$ in mJy beam$^{-1}$ are given for non-detections. 

SM1-A was not detected at 70 $\mu$m in either Spitzer or Herschel data. 
It was marginally detected at 4.5 and 5.8 $\mu$m.
SM1-A is also associated with X-ray emission with Chandra, which is already pointed out by \citet{frie14}.
Source-X was detected only at 219 GHz, 3-mm (Kirk et al. 2017), and X-ray with Chandra. 
The 3 $\sigma$ upper limits at $K_s$, IRAC, MIPS, and {\it Herschel} bands are 
almost identical to those obtained for SM1-A.

Below we summarize the observational results for SM1-A and Source-X. 
The parameters obtained here are summarized in Tables \ref{tab:properties2} (cores) and \ref{tab:outflow} 
(associated CO outflows).

\subsection{Detection of Compact Cores Associated with Chandra Sources}

\begin{figure}
\epsscale{1.0}
\plotone{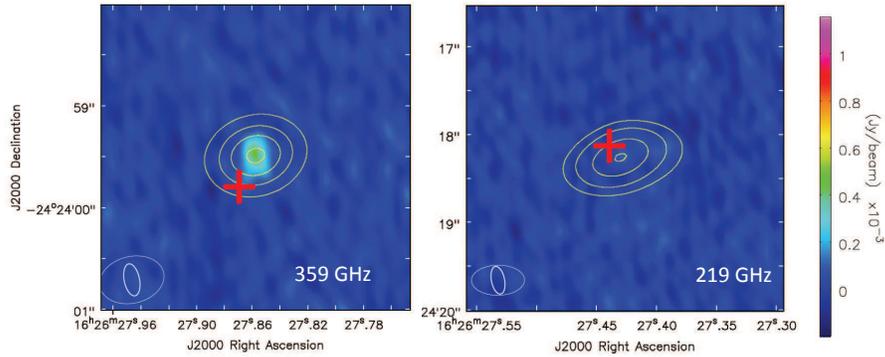}
\caption{ALMA 359$/$219 GHz (contours) and JVLA 41 GHz (color) images. 
({\it left}) SM1-A images at 359 GHz and at 41 GHz. 
({\it right}) Source-X images at 219 GHz and 41 GHz. Color bar for 41 GHz images is shown. 
Contours levels are 20, 40, 60, and 80 {$ \% $} of each peak; the peaks are 245 mJy beam$^{-1}$ 
with {$\sim$} 600 {$\sigma$ } for 359 GHz image, and 21.2 mJy beam$^{-1}$ 
with $\sim$ 80 {$\sigma$ } for 219 GHz image. The 41 GHz peak of SM1-A is 
0.443 $\pm$ 0.038 mJy beam$^{-1}$. Source-X is not detected above 
3 $\sigma$ at 41 GHz  (3 $\sigma$ upper limit is 0.114 mJy beam$^{-1}$). 
Positions of Chandra X-ray sources are obtained from the combined X-ray image,
which are indicated with the red-crosses.
ALMA and JVLA beams are shown as white ellipses; smaller one is for JVLA. 
\label{fig2}}
\end{figure}

In Figure \ref{fig1}, we present the combined 12-m Array and 7-m Array 1.3-mm continuum image 
of Oph A obtained in ALMA Cycle-2.
In the image, we detected two very compact and bright 1.3 mm cores, SM1-A and Source-X which is located  between VLA 1623 and SM1-A, 
together with a number of less bright sources. 
The positions of the two cores and some other young stellar objects are also shown in Figure \ref{fig1}.
We will describe the details of the image in a separate paper (Hara et al. in prep.).
In the present paper, we focus on the two compact, bright cores. 
The JVLA 41 GHz images of the individual cores are shown in Figure \ref{fig2} with contours 
of the ALMA 359 GHz continuum emission. We detect  41 GHz continuum emission toward 
SM1-A, whereas we do not detect 41 GHz continuum emission above a 3 $\sigma$ level 
from Source-X.

Detection of SM1-A with ALMA has already been reported by 
\citet{frie14}. 
SM1-A and Source-X were also identified from the 3-mm  
ALMA Cycle-2 observations \citep{kirk17}. 
However, both SM1-A and Source-X are more prominent in our 1.3-mm image observed with the 
higher angular resolution. 

\begin{figure}
\epsscale{0.5}
\plotone{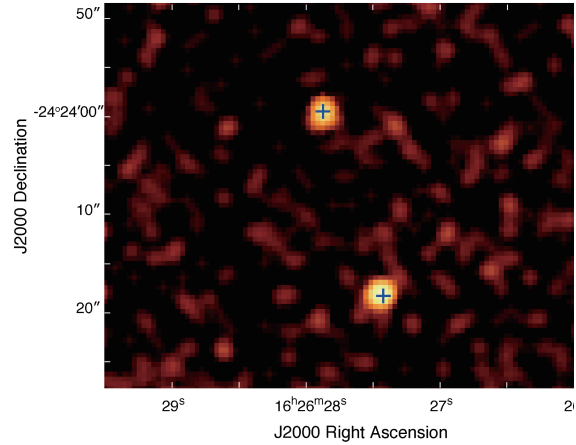}
\caption{
X-ray image around SM1-A and Source-X obtained with Chandra. The data taken in 2000 (Obs ID 637) 
and 2014 (Obs ID 17249) are merged. The crosses indicate the positions of the two sources derived from the ALMA data. 
The image is smoothed with a Gaussian profile setting the width (1$\sigma$) of 3 pixels by using DS9. 
The color shows the number of photon counts per each pixel with a side of 0.5”.  
The color varies with logarithmic scale, as shown in the color bar.
}
\label{fig:xray}
\end{figure}

\begin{figure}
\epsscale{1.0}
\plotone{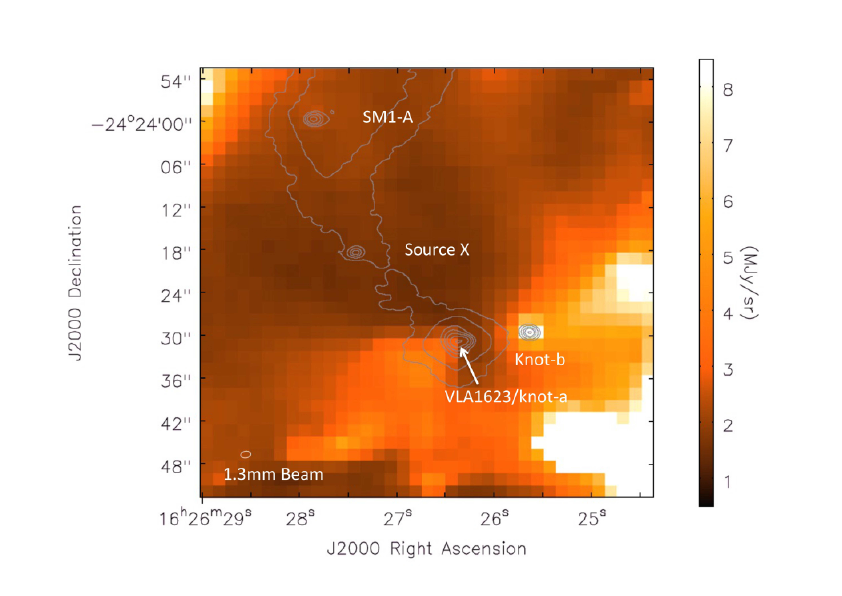}
\caption{1.3 mm (226 GHz) continuum image (contour) taken with the ALMA 12-m and 7-m Arrays superposed on IRAC/{\it Spitzer} 4.5 {$\mu$}m image (color). 
Contour levels are (1, 2, 5, 10, 20, 40) $\times$ 5 mJy beam$^{-1}$ ($\sim$ 20 $\sigma$).  
Color bar for the IRAC image is shown on the right.  
\label{fig4}
}
\end{figure}

The Chandra data merged for the data-set taken in 2000 and that in
2014 revealed that both SM1-A and Source X are significantly detected in
X-ray; the significance levels of SM1-A and Source-X are 13$\sigma$
and 19$\sigma$, respectively, in the 2--10 keV band as shown in Figure \ref{fig:xray}.
The positions of Chandra X-ray sources are shown as red crosses in Figure \ref{fig2}. 
Both of the X-ray detections agree well with the ALMA sources within the 
Chandra positional accuracy of $\sim$ 0$\arcsec$.5 $-$ 1$\arcsec$. 
Analysis of available NIR-FIR/Chandra data unveiled the following nature of both sources: 
\begin{itemize}
\item[(1)] Both sources are very dim in NIR to FIR. SM1-A is marginally 
detected only at IRAC 4.5/5.8 $\mu$m at a $\sim$ 3 $\sigma$ level. 
Source-X is not detected in all NIR-FIR bands.
Although various NIR to FIR data toward 
these two sources have been available, clear signatures of emission from the central 
objects have not 
been previously obtained. In Figure \ref{fig4}, we present 4.5 $\mu$m  
{\it Spitzer} images of the region including SM1-A and Source-X.

\item[(2)] 
Both sources are detected in the Chandra merged map above a
13$\sigma$ level in the 2--10 keV band. As \citet{ima03}
has already reported, source X had
an X-ray flare once with a timescale of a few hours.
This clearly indicates the magnetic
activity on low-mass stellar/sub-stellar core, or a (proto-)BD.

\end{itemize}

Together with the above IR behaviour, these facts indicate ``stellar'' cores 
(i.e., $R_\odot$ –size central sources which have already
experienced the second core collapse) have
already formed at the centers of both sources, 
but the central objects are still extremely young and embedded. 
Hereafter, we call such young objects ``extremely-young Class 0 objects".

\subsection{Detection of Molecular Outflows} 

Another important evidence for protostellar sources to distinguish between prestellar and protostellar cores is the presence of molecular outflows and jets.
In the following we search for high velocity components of molecular outflows from SM1-A and Source-X using the ALMA $^{12}$CO data. 

\subsubsection{SM1-A}

\begin{figure} [h]
\epsscale{1.0}
\plotone{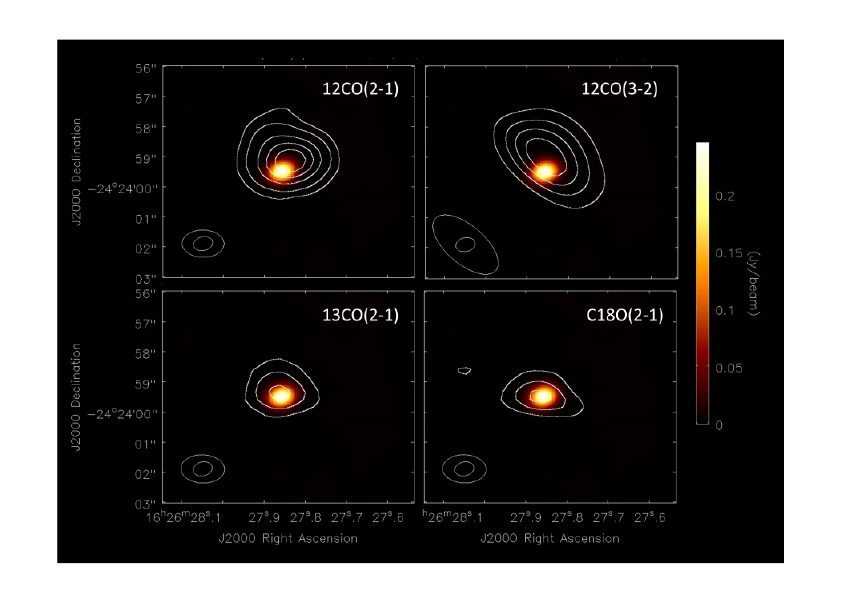}
\caption{Contour maps of blue-shifted emission in 
$^{12}$CO ($J=2-1$), $^{12}$CO ($J=3-2$), $^{13}$CO ($J=2-1$), and  
C$^{18}$O ($J=2-1$) for SM1-A; $V_{\rm lsr}$ = $-$1 to 2 km s$^{-1}$. 
The continuum image at 359 GHz is superposed. 
Contour levels are:  10, 20, 40, 60, 80 {$ \% $} of peak 
( 3.56 $\pm$ 0.094 Jy beam$^{-1}$ km s$^{-1}$) for $^{12}$CO ($J=2-1$);  
20, 40, 60, 80 {$ \% $} of peak (18.5 $\pm$ 0.55 Jy beam$^{-1}$ km s$^{-1}$)  
for $^{12}$CO ($J=3-2$); and 30, 60, 90 {$ \% $} of peaks for $^{13}$CO and C$^{18}$O; 
peaks are 0.81 $\pm$ 0.032 Jy beam$^{-1}$ km s$^{-1}$ and  0.182 $\pm$  0.009 Jy beam$^{-1}$ km s$^{-1}$, respectively. 
Beam sizes are shown as ellipses, with a smaller white ellipse indicating the beam at 359 GHz.}
\label{fig5}
\end{figure}

Very compact $^{12}$CO ($J=3-2$) and $^{12}$CO ($J=2-1$) emission was 
 detected in four channel maps with central velocities of 
 $V_{\rm lsr}$ = $- 0.7$ km s$^{-1}$ to 1.6 km s$^{-1}$ ($\Delta V$ = 0.7 km s$^{-1}$) and 
 $-1$ km s$^{-1}$ to 2 km s$^{-1}$ ($\Delta V =$ 1.0 km s$^{-1}$), respectively. 
 The total integrated intensity maps are shown in Figure \ref{fig5} together 
 with those in $^{13}$CO ($J=2-1$) and C$^{18}$O($J=2-1$), which were integrated 
 over $V_{\rm lsr}$ =  $\sim$ 0.5 km s$^{-1}$ to 2.0 km s$^{-1}$. 
The emission is point-like and the peaks are very close to the continuum 
 source position. For example,  both $^{12}$CO peaks are located 0.$\arcsec$7 north of SM1-A. 
 Since the systemic velocity of the SM1 core is known to be $\sim$ 3.8 km s$^{-1}$ 
from observations of optically thin molecular lines such as N$_2$H$^+$ ($J=1-0$) 
\citep[e.g.,][]{difrancesco04}, 
  the CO emission is blue-shifted by 2 $-$ 4 km s$^{-1}$. 
 There is no other significant $^{12}$CO emission in the ALMA and SMA maps 
 in the 20$\arcsec$ $-$ 30$\arcsec$ region around SM1-A although redshifted 
 CO emission located $\sim$20$\arcsec$ NE of SM1-A is seen in the SMA maps 
 (near the edge of the FOV). This emission likely corresponds to the red-shifted 
 part of an outflow from a NIR source, GY 30, found by a past CO outflow search 
 \citep {kama03}. The total integrated intensities of the blueshifted $^{12}$CO ($J=3-2$) and  $^{12}$CO ($J=2-1$) emission 
 are 26 $\pm$ 0.55 Jy km s$^{-1}$ 
 and 7.34 $\pm$ 0.28 Jy km s$^{-1}$, as summarized in Table \ref{tab:outflow}. The beam 
 de-convolved size and the brightness temperature  were measured to 
 be $\sim$ 1.$\arcsec$2 $\times$ 0.$\arcsec$6 and $\sim$ 60 $-$ 100 K, respectively. 
 The simple interpretation of the blue-shifted CO emission is a very compact and 
 low velocity outflow from the stellar core. 
 The lack of a redshifted counter part is puzzling and 
 might be due to the asymmetry in velocity structure or distribution 
 of the molecular outflow itself. In addition, surrounding/ambient molecular 
material  that potentially could have some influence (absorption) on the outflow, 
 especially around the systemic velocity. The mass of the outflow 
 was estimated to be between 1.3 $\times$ 10$^{-4}$ $M_\odot$ and 1.3 $\times$ 10$^{-5}$ $M_\odot$ where the lower and upper boundaries 
 are for the optically thin and thick wing emission cases, respectively. 
 (For the wing, we obtained $\tau\sim 10$ based on unpublished 
 single-dish $^{12}$CO ($J=3-2$) and $^{13}$CO ($J=3-2$) data.) 
 Here, we assumed LTE condition, the CO fractional abundance of $10^{-4}$, and the excitation temperature of 20 K.
The dynamical time is roughly $\sim$ 430 yr assuming the outflow length is roughly 
 $\sim 270/\cos i$ au  and the velocity is $3/\sin i$ km s$^{-1}$
($i$ is measured from the plane of sky) and $i$ = 45$\arcdeg$. 
Note that even if $i$ = 75$\arcdeg$, the time is still as short as $\sim$ 550 yr. 
This outflow would be one of the most compact molecular outflows known. 
The inferred mass loss rate, 10$^{-6} - 10^{-7}$ $M_\odot$ yr$^{-1}$, 
is consistent with that expected from a typical low-mass protostar.

\subsubsection{Source-X}

\begin{figure} [h]
\epsscale{1.0}
\plotone{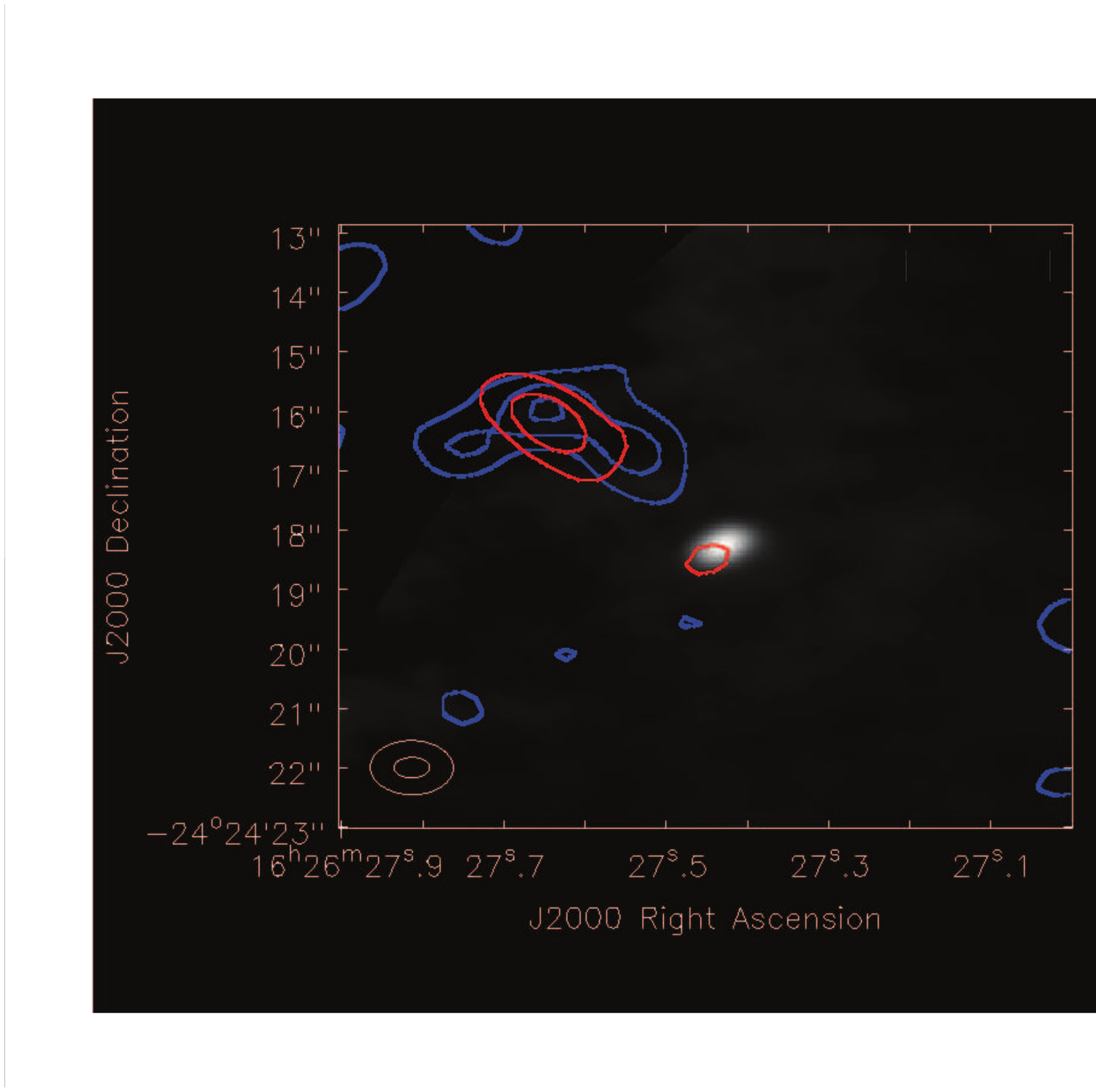}
\caption{Images of blue-shifted (blue line) and red-shifted (red line) 
emission in $^{12}$CO ($J=2-1$) for Source-X; $V_{\rm lsr}$ = 2 km s$^{-1}$ and 7 km s$^{-1}$ 
with a velocity width of 1 km s$^{-1}$ for each. 
The continuum image (color) at 219 GHz is superposed. 
Contour levels are 3, 5, and 7 $\sigma$ (1  $\sigma$ = 0.04 Jy beam$^{-1}$ km s$^{-1}$) for each. Flux density scale is shown on the right. 
Beam sizes are shown as ellipses, with white ellipse indicating the beam at 219 GHz. }
\label{fig6}
\end{figure}

The $^{12}$CO ($J=2-1$) integrated intensity contours superimposed on the 219 GHz continuum image
are shown in Figure \ref{fig6}.
We detected faint blue- and red-shifted emission in 
$^{12}$CO ($J=2-1$) to the $\sim$ 4$\arcsec$ north-east of Source-X, 
roughly perpendicular to the elongation of dust emission at 219 GHz.
Although there is no southwestern counterpart  in the $^{12}$CO ($J=2-1$) emission components,
a possible interpretation of these components is that they are the high-velocity 
outflow components from Source-X, and the outflow axis is almost parallel to the plane of sky.
If this emission is indeed a low-velocity outflow, then we could compute the outflow parameters of Source-X which are listed in Table \ref{tab:outflow}.
The total integrated intensity of the $^{12}$CO ($J=2-1$) emission is estimated to be 1.14 $\pm$ 0.097 Jy km s$^{-1}$, which is
smaller than that of SM1-A.
The outflow mass is only $10^{-5}$ $M_\odot$ and the mass loss rate is
evaluated to be about $10^{-8}$ $M_\odot$ yr$^{-1}$.

\subsection{SED} 

\begin{figure} [h]
\plotone{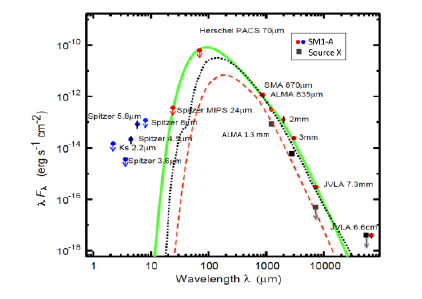}
\caption{Observed SEDs for SM1-A and Source-X together with two model
 SEDs for SM1-A and one model for Source-X. Filled red circles (SM1-A) 
 and a black square (Source-X) with error bars of $\pm$ 1 $\sigma$ are detections 
 above 3 $\sigma$ and filled blue circles and black squares with arrows are 
 3 $\sigma$ upper limits at e.g., Ks, IRAC 3.6/8 $\mu$m, Spitzer 24 $\mu$m, and 
 Herschel 70 $\mu$m. We used photometry data summarized in Table 2. 
 The models are the calculated SED for a rotating first core model (dotted line) 
 and a uniform temperature core model (light green line for SM1-A, or 
 red broken line for Source-X). The first core SED is obtained for the aperture 
  $r <$ 120 au (2$"$ in diameter). Recently, \citet{kirk17} detected Source-X in 3 mm in ALMA Cycle-2 observations.
  Their source No. 10 corresponds to Source-X whose flux density is 7.56 mJy at 3 mm.}
  \label{fig7}
\end{figure}

Figure \ref{fig7} shows the SEDs of SM1-A and Source-X. 
There are several detections of cm to submm emission toward SM1-A, 
but only two detections at 219 GHz and 226 GHz toward Source-X. 
The 3 $\sigma$ upper limits to NIR to FIR flux densities in Source-X 
are the same as those in SM1-A. 
The tentative detections in 4.5 $\mu$m and 5.8 $\mu$m are 
only 3 $\sigma$ upper limits in SM1-A.

We compare the SED of SM1-A with the theoretical model of the first core  in Figure \ref{fig7}.
The SED of SM1-A looks similar to that of the first core, indicating that
SM1-A is in the extremely early phase close to the first core formation phase.

\subsection{Source Size Analysis} 

\begin{figure} [h]
\epsscale{1.0}
\plotone{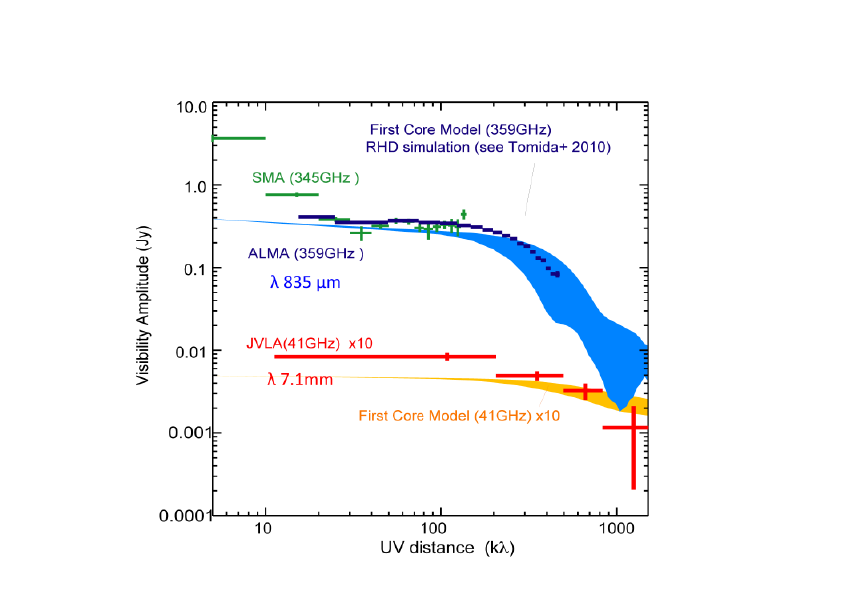}
\caption{ALMA, SMA, and JVLA visibility amplitudes for  SM1-A. 
The SMA visibility amplitudes at 870 $\mu$m (filled green square), ALMA 835 $\mu$m 
(filled blue square), and JVLA 7.3 mm (filled red square) are plotted as a function of $uv$ distance, together with the rotating first core models from RHD calculations \citep{saigo11, tomi10} 
for $i = 60^\circ$ 
in the late evolutionary phase.  Light blue and orange lines are for an initial cloud mass of the 
0.1 $M_\odot$ giving a better fit to the data  for the 1 $M_\odot$ model. 
The error in the vertical scale is 1 $\sigma$  standard deviation estimated in each bin 
of the $uv$ distance shown as a horizontal bar. 
The same models were used for calculating the model SEDs shown in Figure 6. 
}
\label{fig8}
\end{figure}

The source size can be derived from the beam-deconvolved size
of the millimeter and submillimeter continuum emission. 
For SM1-A and Source-X, the beam-deconvolved sizes derived 
from the 359/219 GHz images (Figure \ref{fig2}) are 0$\farcs$3 $\times$ 0$\farcs$3
and 0$\farcs$6 $\times$ 0$\farcs$3, respectively. 
The beam deconvolved size is also derived to be  0.$"$2 $\times$ 0.$"$2 from the 41GHz image for SM1-A. 
Another method to determine size is visibility fitting 
assuming that the source intensity has 
an axisymmetric Gaussian distribution. For the assumption,
we adopt the following relation;
\begin{eqnarray}
\Bigl( \frac { UV_{1/2} } { 100 \:k\lambda}  \Bigr) \cdot 
\Bigl( \frac {  \Theta_{\rm FWHM} } { 1 \: \rm arcsec}  \Bigr)  &=& 
\frac {4\: \ln 2} {2\pi} \cdot  \frac {360\cdot 3600} {2\pi\cdot 10^5}     \nonumber \\
\nonumber\\
  &=& 0.9111 
\end{eqnarray}
where $UV_{1/2}$ is the $uv$ distance where the visibility amplitude 
is a half of the peak, and  $\Theta_{\rm FWHM} $  is the source size (FWHM) 
in the image plane. 
In Figure \ref{fig8}, we show the visibility plots of SM1-A.
We also compared the 359 GHz and 41 GHz visibility distributions with those expected for the first core model (Tomida et al. 2010).
The 3D radiative hydrodynamics (RHD) calculation started from critical Bonnor-Ebert (BE)-like spheres as an initial condition, which have 
$T=$ 10 K, and densities increased by a factor of 1.6 to make them unstable. The calculations were done for 
initial clouds with masses of $M=$ 0.1 and 1.0 $M_\odot$. 
The SEDs and visibility distributions were calculated for inclination angles of 60$^\circ$,
and those for the 0.1 M$_\odot$ model giving the better fits to the observations are shown in Figures 7 and 8.

The obtained source diameters from the visibility amplitude fitting of SM1-A 
are 0.3$''$ (= 41 au) and 0.17$''$ (= 23 au) for 359 GHz and 41 GHz, respectively. 
The source diameters derived from the visibility analysis agree well with 
the beam-deconvolved sizes in the image domain
at the same frequencies.  There is a large discrepancy between the diameters derived from 359 GHz and 41 GHz, which 
is likely to originate from the different  optical depths (The 359 GHz emission of SM1-A is likely to be optically thick, whereas 
the 41 GHz emission is optically thin. See Section 4.1.1 for more detail.)
The source size is a half that of  \citet{frie14}, who derived the effective radius of 42 au,
where we corrected the difference in the assumed distance.

For Source-X, we do not have enough visibility data to do the same analysis as done for
SM1-A. We adopt the  geometric mean of the beam-deconvoled size, 0.4$\arcsec$ (= 55 au),
as the source size of  Source-X.  
\citet{kirk17} measured the deconvolved  size of $1\arcsec.08 \times 0\arcsec.36$ for Source-X,
which is slightly larger than our estimate.

\subsection{X-ray Properties; $N_H$ and $L_X$}

From the merged data, we made averaged source spectra, the response
files, and the background files, for SM1-A and Source X, using the
CIAO software ``combine\_spectra". Both of the source regions
are a circle with radius of 3 arcsec, centered at the respective sources.
A common background region was
used for both sources, taken from a source-free region 
in a circle with radius of 37 arcsec, centered at 
 ($\alpha _{J2000.0}$, $\delta _{J2000.0}$)
= (16$^{\mathrm h}$26$^{\mathrm m}$29\fs390, $-$24\arcdeg24\arcmin52\farcs94).
The background region is located at the same CCD chip which both sources are on.

We fitted the spectra using the thin thermal plasma model (the Astrophysical Plasma Emission Code, Smith et al. 2001)
along with the photoelectric absorption model (WABS: Balucinska-Church \& McGammon 1992).
In the photoelectric absorption model, we adopted the Wisconsin cross-sections (Morrison \& McCammon 1983)
and the \citet{anders82} relative abundances.
The metal abundances and the plasma temperatures were fixed to 0.3 solar and 5
keV, respectively, based on the previously derived values for Young
Stellar Objects (e.g. Imanishi et al. 2001).  The best-fit values of the absorbing columns and the
absorption-corrected X-ray luminosities in the 0.5--10 keV band, with
the errors for them in 90\% confidence level, are 
$(N_H, L_X) =$ $(3.0^{+2.2} _{-1.2} $ $\times 10^{23}$ cm$^{-2}$, $0.75^{+1.0}_{-0.5} \times10^{29}$ erg s$^{-1}$) and 
$(3.4^{+3.7} _{-2.1} $ $\times 10^{23}$ cm$^{-2}$, $0.78^{+2.1}_{-0.6} \times10^{29}$ erg s$^{-1}$)  for SM1-A and Source-X, respectively.

\section{Discussion}

\subsection{SED Fit and Derivation of Physical Properties} 

\subsubsection{SM1-A}

The most remarkable new finding is that SM1-A is extremely dense and 
has an optically thick surface at 359 GHz with a radius of r $\sim$21 au 
as derived below. Firstly, the SED fit with a uniform dust temperature 
($T_{\rm d}$) source model with a radius 
0.$''$15 ( = 21 au) was taken to derive $T_{\rm d}$, 
the optical depth at 7.3 mm (41 GHz) $\tau _{41}$, and the dust opacity index $\beta$. 
The 7.3-mm continuum emission is assumed to arise from the thermal dust emission, 
without  free-free contamination.
These parameters allow us to infer the luminosity, mass, and density 
of the core. We fit the observed SED using the following equations to estimate $T_d$,

\begin{equation}
S_{\nu} =  \bigl(  B_{\nu} (T_{\rm d})  - B_{\nu} (T_{\rm bg} ) \bigr) 
\bigl[ 1-\exp \bigl(- \tau_{\nu} \bigr) \bigr] \Omega , 
\end{equation}
where  $T_{\rm bg}$ is the temperature of the cosmic background radiation and set to 2.7 K.
$\Omega$ is the source solid angle for SM1-A,  
which is assumed to be 2.4 $\times$ 10$^{-12}$ $sr$ 
(corresponding to a source diameter of 41 au) 
at both 41 GHz and 359 GHz for simplicity.   
The function  $B_\nu (T)$ is the  Planck function and the opacity $\kappa_\nu$ is expressed as  
\begin{equation}
\kappa_{\nu} =  0.1 \Bigl( \frac {\nu } { 10^{12}\: {\rm Hz}}    \Bigr)^{\beta} \;\;\; {\rm cm}^2 \; {\rm g}^{-1} .
\end{equation}

The fit to its SED, 
i.e, the obtained flux densities at 7.3 mm and 835 $\mu$m (359 GHz) together 
with the 24 $\mu$m and 70 $\mu$m 3 $\sigma$ upper limits, 
gives us $T_{\rm d} = 40$ K, $\beta$ = 1.5 $-$ 2, and $\tau _{41} = $ 0.18. 
The optical depth at 835 $\mu$m is expressed as $\tau_{359}$ = (7.3/0.835)$^{\beta}$ $\times$ $\tau_{41}$, hence $\tau_{359}$ = 5 $-$ 14. 
This value indicates that SM1-A has an optically thick surface at submillimeter wavelengths, 
i.e., a submillimeter photosphere, at $T_{\rm d}$ = 40 K. 
On the other hand, the optical depth at 41 GHz is much smaller than unity, i.e., SM1-A is optically thin.
The object presumably has the centrally-condensed density distribution.  
The optically-thin 41 GHz emission can trace the inner denser region which
cannot be seen at the optically-thick 359 GHz.
This would be the reason why the estimated source size at 41 GHz is smaller 
than that at 359 GHz.

Secondly, 
we estimate the luminosity and density of the core, where we assume a uniform-density 
spherical core with the opaque surface at $r =$ 21 au and a temperature of 40 K. 
The luminosity of SM1-A  calculated so is  0.035 $L_\odot$, directly 
using 
\begin{equation}
L = 4\pi r^2\sigma_{\rm SB} T_{\rm d}^4 \ ,
\end{equation}
where $\sigma_{\rm SB}$ is the Stefan-Boltzmann constant. 
The opacity-corrected mass is given as
\begin{equation}
M = \frac {\tau_{\nu} } {1-\exp(-\tau_{\nu}) } \: \frac {S_{\nu} D^2} {\kappa_{\nu} B_{\nu} (T_{\rm d}) } \ ,
\end{equation}
and we estimate to be 0.054$-$0.27 $M_\odot$ for $T_{\rm d} =$ 40 K,  
$\beta=$ 1.5 $-$ 2, and $D = 137$ pc. 
This is in good agreement with that of \citet{frie14}.
Assuming spherical symmetry,  the mean number density and volume density are estimated to be 
$n= (2.2 - 8.4) \times 10^{11}$ cm$^{-3}$ and  
$\rho = (5.2 - 33) \times 10^{-13}$ g cm$^{-3}$, respectively,
which meet those expected for a first core \citep{masu98,bate02,saigo08,tomi10,tomi13}. 
In addition, the luminosity, 0.035 $L_\odot$,  is mostly in the range of that predicted for a first core
($L_{\rm int} < $ 0.06 $L_\odot$). 
For comparison, we  list the physical quantities of the first core predicted from
numerical simulations in Table \ref{tab:properties2} \citep{lar69,masu98,saigo08}.
This similarity in physical quantities between SM1-A and the first core models indicates that 
SM1-A  is extremely young object. 
Taking into account the fact that molecular outflows and X-ray emission are detected,
SM1-A is likely to be an extremely-young proto-brown dwarf or protostar which already has
a second core inside (i.e., an extremely-young Class 0 object).

\subsubsection{Source-X}

Source-X is lacking data to constrain its physical nature due to its limited  detection
 at 100/219/226 GHz. Instead of modelling, we assume 
 $T_{\rm d}$ = 20 K  ( $T_b \sim$ 12 K is obtained for the source at 218 GHz, 
 and 
 $T_{\rm d}$ should be higher than that if the emission at 
 218 GHz is not optically thick.)  In addition, we  assume $\beta=1.5-2$ 
 since the spectral index between 218 GHz and 41 GHz is larger than 3.4. 
 These assumptions and the observed properties yield a  mass range 
 of $M=0.018-0.039$ $M_\odot$ and a number density of $n\simeq 6.0 \times 10^{10}$ cm$^{-3}$. 
 \citet{kirk17} derived  the mass of 0.071 $M_\odot$ in 3 mm, somewhat larger than our estimation,
 but it is consistent with each other, taking into account the effects of the different parameters.
 These values are also consistent with those predicted by the first core model.
 This indicates that Source-X is in the similar evolutionary phase as SM1-A.
 
The derived physical parameters of Source-X are listed in Table \ref{tab:outflow}.

\subsection{Comparison of X-ray Properties among YSOs in Oph}

\begin{figure} [h]
\epsscale{0.9}
\plotone{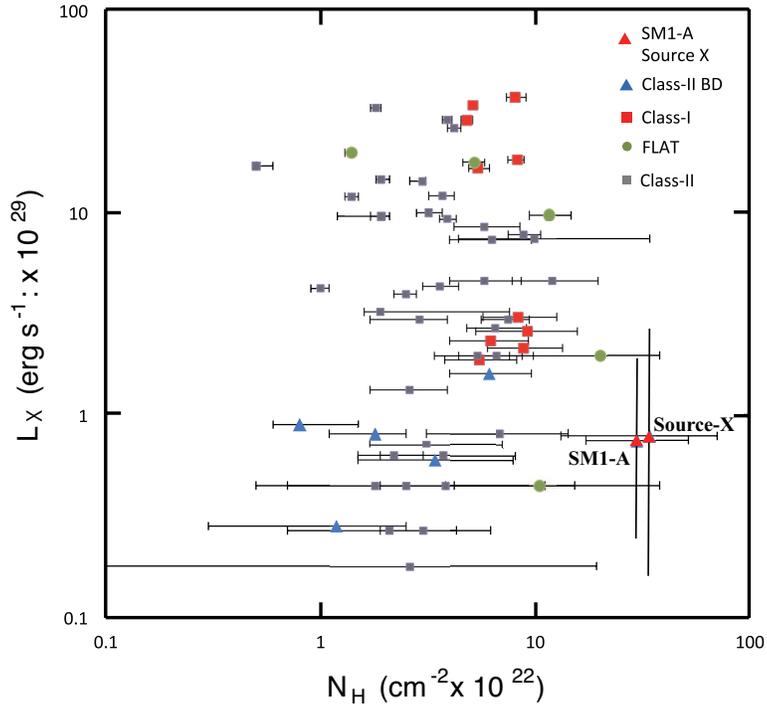}
\caption{Column density ($N_H$) vs. absorption-corrected X-ray luminosity for YSOs.
Data except for SM1-A and Source-X were taken from \citet{ima03}.
$N_H$ and $L_X$ for SM1-A and Source-X were esstimated from the combined Candra data 
(see the text).
Classifications of YSOs are based on the c2d catalogue of Evans et al. (2008).
The Class II BD sample is based on \citet{alves13}.
}
\label{fig:Nh_Lx}
\label{fig9}
\end{figure}

The absorbing column densities ($N_H$) and absorption-corrected X-ray luminosities  ($L_X$) for SM1-A and Source-X are plotted in Figure \ref{fig:Nh_Lx}.
The values for YSOs in Oph reported
by \citet{ima03} are also plotted, after conversion of the
X-ray luminosity using updated distance of 137 pc. The errors for the
X-ray luminosities of the YSOs are not
indicated, because \citet{ima03} did not provide the errors of the X-ray luminosities.
Interestingly, the derived X-ray luminosities of
SM1-A and source X are almost the same level of those for Class-II
type BDs, although they are not inconsistent with the other types
(e.g. Class-II  non-BDs). 

The column density of
absorbing material derived from the X-ray emission is
$\sim 3\times10^{23}$ cm$^{-2}$, the highest among Class-I 
Chandra sources in the Oph region.
This is more evidence that the sources are extremely young Class 0 objects.
On the other hand, column densities are estimated to be an order of $10^{25-26}$ cm$^{-2}$ from 
the obtained masses or number densities assuming the cores are spherically symmetric and 
uniform in density: e.g., a core density of $\sim 10^{11-12}$ cm$^{-3}$ and a core radius 
of $\sim 20$ au ($3 \times$  10$^{14}$ cm) produce the column density to the central stellar core, 
$\sim 3 \times$ 10$^{25-26}$ cm$^{-2}$.  This discrepancy between X-ray and mm/sub-mm estimates will be reconciled 
if the cores have not spherically symmetric but disk-like structures where column densities decrease rapidly 
as viewing angles changing from edge-on to pole-on. In fact, Source-X seems to have a disk-like structure perpendicular 
to the outflow since the beam decomvolved size of the 219 GHz image is 
$0$\farcs$6 \times 0$\farcs$3$ with PA=124 deg as noted in Section 3.5.  
According to theoretical studies (e.g., Saigo et al. 2008), a centrifugally supported disk as a remnant of the first core 
remains even after the stellar core formation in a rotating core. The extremely dense cores in SM1-A and Source-X 
would be such remnants of the first cores surviving in extremely-young  Class-0, which can be detected in X-ray 
due to preferable viewing angles with less absorbing column  densities.

\subsection{SM1-A and Source-X are proto-BDs?}

As we showed above, we discovered two extremely-young Class 0 objects with substellar masses.  The fates of these low-mass objects remain uncertain.
Taking into account the masses derived from the dust continuum emission, they can potentially evolve into brown dwarfs unless they gain significant masses from the surroundings.
There are at least two widely-discussed scenarios for brown dwarf formation (Machida et al. 2009, Basu \& Vorobyov 2012).
One is the gravitational contraction of a very low-mass core. 
This scenario considers that brown dwarfs form similarly to low-mass stars.  Starting from a spherical magnetized core with 0.22 M$_\odot$,
Machida et al. (2009) demonstrated that brown dwarf can form from such a low-mass dense core.
Another scenario is the dynamical ejection from multiple systems (e.g., Reipurth \& Clarke 2001) or circumstellar disks (e.g., Basu \& Vorobyov 2012).
Here, we briefly discuss the possibility that these objects evolve into brown dwarfs on the basis of the two scenarios.

\subsubsection{Comparison with Machida et al. (2009)}

\citet{machi09} performed 3D MHD simulations of  the brown dwarf formation.
Their initial condition was a critical Bonner-Ebert sphere of masses of 0.22 $M_\odot$.
For SM1-A and Source-X, the dynamical times of the outflows are estimated to be $\sim 500$ yrs.
In the model by \citet{machi09}, the outflow mass and mass loss rate are derived to be $\sim 2 \times 10^{-3}$ $M_\odot$ and
$\sim 4 \times 10^{-6}$ $M_\odot$ yr$^{-1}$, respectively, at the evolution time of $\sim$ 500 yrs from the first core formation epoch.
The luminosity is estimated to be $0.2-1.0$ $L_\odot$ with a radius of 2 $R_\odot$.
This is somewhat smaller than the values derived from the observations.  
The outflow mass and mass loss rate are significantly smaller than those obtained from the observations.
If SM1-A and Source-X formed from compact dense cores, their evolution seems to be consistent with this numerical simulations.

\subsection{Ejection from VLA1623A protostellar binary?}

According to the dynamical time of 
the molecular outflow detected, the ages of two sources are likely to be 500-1000 yrs 
or less after the central stellar core formation. Such very young sources should be 
very rare even in the Oph star-forming region, hence it will be very unusual that two are 
independently formed and located closely in the small $\sim$5000 au region. 
It should be noted that similar types of sources are not detected with the similar ALMA 149-pointing imaging in the Oph-B2 region (Kamazaki et al 2018). 
One possible common origin of the two sources would be ejections from the VLA1623 region.

Reipurth \& Clarke (2001) proposed that a possible observational test of the ejection 
scenario is brown dwarf searched in the vicinity of Class-0 sources with an age of $\sim10^{4-5}$ years. 
If stellar embryos or first core like dense gas cores are ejected from the Class-0 object in the phase 
of disk/envelope fragmentation, one might expect to detect one or more (proto) brown dwarfs 
around the Class 0  object. The apparent separations from the Class-0 object VLA1623A to SM1-A and Source-X
 are $\sim$5000 au and 2600 au, respectively. The two sources, SM1-A and Source-X, 
are located at PA= 30 to 50 deg. measured from  VLA1623A, and mostly 
aligned with the orbital plane of the disk-like envelope seen in C$^{18}$O (PA= 32 deg; Murillo et al. 2013). 
If we assume their ejection velocity to be roughly 3 km s$^{-1}$, a factor of $\sim$1.5 higher than the rotating 
velocity of the innermost in the envelope, 2 km s$^{-1}$, and also assume the travels are along the 
plane of sky, the dynamical times to travel to the current locations are estimated to be 8000 yrs and 4000 yrs for 
SM1-A and Source-X, respectively. For that two are moving at 60 deg to the plane of sky, 
the timescale is doubled but still roughly consistent with the age of the Class-0 object.

\begin{figure} [h]
\epsscale{0.9}
\plotone{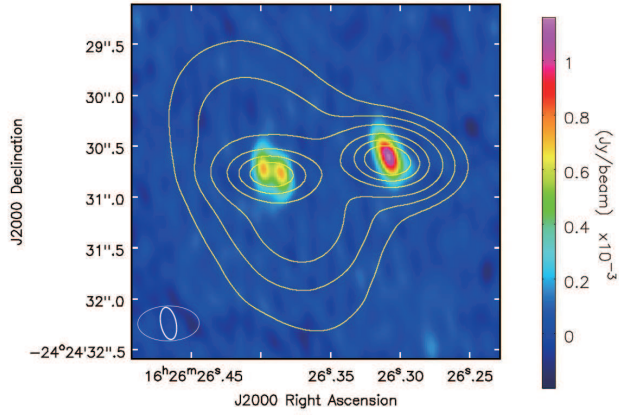}
\caption{JVLA 41 GHz image (color) of VLA1623A and VLA1623B (Knot-a), 
superposed on the ALMA 219 GHz image (contour). 
The image size is 3$"$.5 $\times$ 3$"$.5.  The color scale and beam sizes are the same as in Figure 2. 
The Contour levels are 5, 10, 20, 40, 60, and 80 \% of the peak; the peak is 91.7 Jy/beam 
with $\sim$ 1090$\sigma$. 
}
\label{fig:vla}
\end{figure}

 Unexpectedly, VLA1623A was resolved to two sources separated by 0.2$\arcsec$ ($\sim$25 au) 
 with the recent ALMA observations (Harris et al. 2018) and our JVLA 7mm observations 
 (see Figure \ref{fig:vla}) suggesting an equal-mass ($\sim$ 0.05$-$0.1 M$_\odot$) protostellar binary. 
 The disk-like structure in C$^{18}$O is likely to be 
 common envelope of the binary. The binary would be evidence for disk fragmentation. 
 Each mass of the binary seems to be larger than SM1-A and Source-X, hence the 
 binary could eject the 3-rd less mass object in the system (e.g., Reipurth \& Clarke, 2001). 
These observations seem to be consistent with the ejection scenario, especially hybrid 
scenario (Basu \& Vorobyov 2012); first core like dense clumps are ejected, and the 
ejected clumps formed very low mass stars afterward. 
It is noted that the discrepancy between the traveling times of SM1-A and Source-X 
and the dynamical timescales of the outflow would be reconciled with the hybrid 
scenario, i.e., recent ($\sim$1000 yrs) second collapses in ejected first core like clumps 
triggered outflow and X-ray activity. Furthermore, Tomida et al. (2011) showed with 
radiation hydrostatic simulations that first cores formed in very low-mass cores lives 
longer than $10^4$ yrs. If ejected cores evolve similarly such first cores, they can 
travel ~ 6000 au at $v=$3 km s$^{-1}$ in $10^4$ yr.  

However, further observations should be required to pursue the ejection scenario 
more quantitatively and carefully. Some of clumps may be tidally disrupted during ejection 
and disperse according to the simulation (e.g., Basu \& Vorobyov 2012, Stamatellos \& Whitworth 2009). 
Such failed cores may be found around VLA1623A, and tidally formed arms as fossil of interaction 
may be also detected even around SM1-A and Source X in the further more sensitive observations.
Since the eastern area of VLA 1623 has no significant dense gas, the other ejected objects might be found from
future high-angular resolution, high sensitivity observations.

\section{Summary}

Based on SMA, ALMA, JVLA, and {\it Spitzer} data, we have discovered two candidates of forming brown dwarfs or low-mass
stars in the Ophiuchus A region.  
Detections of  small outflow lobes and  X-ray emission imply that they are in the
extremely early formation phases of protostars or proto-BDs.
The similarity in the physical nature to the FHSCs indicates that they just passed the FHSC formation phase.
 
We summarize the primary results of the present paper as follows: 
\begin{itemize}
\item[1.] We analyzed submillimeter, infrared, and X-ray data of two bright and 
compact dust continuum sources in Oph A, and compiled their SEDs.
One, SM1-A, is located in the densest part of the Oph A ridge, and previously named SM1. 
We call the core SM1-A. The other, Source-X, is located south of SM1-A.  
\item[2.] Both objects are mostly invisible in infrared, but are seen in X-ray, 
with time-variability in Source-X. 
This detection indicates that these objects have already experienced second core collapse.
\item[3.] We detected possible CO outflow lobes for both cores.
For SM1-A, we detected high velocity blue-shifted components in $^{12}$CO ($J=3-2$),
$^{12}$CO ($J=2-1$), $^{13}$CO ($J=2-1$), and C$^{18}$O ($J=2-1$). 
For Source-X, we detected  faint blueshifted and redshifted lobes in $^{12}$CO ($J=2-1$).
The dynamical timescales of the outflows was estimated to be  several hundred years for both sources.
\item[4.] From the SED fits, we derived masses of 0.028$-$0.14 $M_\odot$ and 0.014$-$0.03 $M_\odot$
for SM1-A and Source-X, respectively.  The number densities are  estimated to be $\sim 4.6 \times 10^{11}$ cm$^{-3}$
and $6.0\times 10^{10}$ cm$^{-3}$, for SM1-A and Source-X, respectively. 
The masses of these objects are substellar ($M \lesssim 0.08$ $M_\odot$).  
These values are consistent with those predicted by the first core model.
Taking into account the fact that these sources show the protostellar  signatures, i.e., outflows and X-ray emission,  
these objects presumably passed only several hundred years from the the first hydrostatic core formation. We call such object ``extremely-young Class 0 objects.
\item[5.] From their SEDs, X-ray activity, and CO outflow lobes, 
we speculate that two cores are proto-brown dwarfs if they gain more mass only from remnants of the first cores
or will evolve into protostars after they gain more masses from  surroundings. 
For the latter case, these two objects {might be} stellar seeds in the competitive accretion scenario.
\end{itemize}

\acknowledgments

We are grateful to Ken Yabuki and Yumiko Nakamura for their help on the data reduction and analysis of the Chandra X-ray data.
We thank to Kohji Tomisaka, Yuri Aikawa, Tomoaki Matsumoto, Kazuyuki Omukai for valuable comments and discussion.
We would also like to thank to Sergio A. Dzib for providing us with the VLA data.
We would also like to thank the referee for his or her valuable comments and suggestions which improved the paper greatly
Data analysis was in part carried out on the open use data analysis
computer system at the Astronomy Data Center, ADC, of the National
Astronomical Observatory of Japan.
This work was financially supported by JSPS KAKENHI Grant Numbers JP17H02863 (FN),
JP16H07086 and JP18K03703 (ST), and  JP17K05392 (YT), 
and NAOJ ALMA Scientific Research Grant Number 2017-04A.
YS received support from the ANR (project NIKA2SKY, grant agreement ANR-15-CE31-0017).
NH acknowledges a grant from the Ministry of Science and Technology (MoST) of Taiwan (MosT 107-2119-M-001-029).
This paper makes use of the following ALMA data:
ADS/JAO.ALMA\#2011.0.00396.S, ADS/JAO.ALMA\#2013.1.01004.S and ADS/JAO.ALMA\#2013.1.00839.S.
ALMA is a partnership of ESO (representing
its member states), NSF (USA) and NINS (Japan), together with NRC (Canada)
and NSC and ASIAA (Taiwan) and KASI (Republic of Korea), in
cooperation with the Republic of Chile. The Joint ALMA Observatory is
operated by ESO, AUI/NRAO and NAOJ.

\facility{ALMA, SMA, JVLA}

\clearpage

\clearpage

\clearpage

\begin{deluxetable}{lllllllll}
\tablecaption{Summary of Interferometric Observations\label{tab:interferometer}}
\tablehead{
\colhead{Freq.} & \colhead{Array} & \colhead{date} & \colhead{RA (J2000)} & \colhead{decl (J2000)} &
\colhead{FOV} & \colhead{Beam Size (PA)} & \colhead{rms$^a$} & \colhead{Comment}\\
\colhead{ [GHz]} & 　& 　& \colhead{[h:m:s]} & \colhead{[\arcdeg: \arcmin: \arcsec]} & \colhead{[\arcsec] } &\colhead{arcsec$\times$arcsec ($^\circ$)]} & \colhead{[mJy] }  & 
}
\startdata
41 & JVLA &  2012 Sep 03 & 16:26:27.4 & $-$24:24:08   &  \colhead{ 73} & \colhead{ 0.32 $\times$ 0.15 ($+$11.4)} & 0.038  &    \\
219 & ALMA & 2014 Aug 17 & 16:26:26.39  & $-$24:24:30.688    &   \colhead{29} &  \colhead{ 0.6 $\times$ 0.34 ($-$89.27)} & 0.084  & Cycle-2$^c$    \\
226 & ALMA & 2015 Mar 01 & 16:26:27.6 & $-$24:23:55.0   &   180 $\times$ 120$^b$ & \colhead{ 1.4 $\times$ 0.92 ($-$81.1)} & 0.024 & Cycle-2$^d$   \\
345 & SMA & 2007 July 29 & 16:26:27.6  & $-$24:23:55.0    &  \colhead{ 36} &  2.6 $\times$ 1.26 ($+$49.3) & 10.5  &    $^e$ \\
359.2 & ALMA & 2012 Aug 24 & 16:26:27.83  & $-$24:23:59.2   & \colhead{  17.3 }& \colhead{ 0.64 $\times$ 0.46 ($-$76.7)} & 0.39  &  Cycle-0$^f$  \\
372.4 & ALMA & 2012 July 02 & 16:26:27.83 &  $-$24:23:59.2   &  \colhead{ 16.7 } & \colhead{ 0.59 $\times$ 0.42 ($-$75)} & 1.15  &  Cycle-0$^f$    
\enddata
\tablenotetext{a}{rms noise measured at the central part of each image}
\tablenotetext{b}{Obtained with 150 pointings of 12-m and 7-m Arrays, the ALMA data ID is ADS/JAO.ALMA\#2013.1.00839.S.}
\tablenotetext{c}{The ALMA data ID is ADS/JAO.ALMA\#2013.1.01004.S.}
\tablenotetext{d}{\citet{kama18}}
\tablenotetext{e}{Nakamura et al. (2012)}
\tablenotetext{f}{Friesen et al. (2014), the ALMA data ID is ADS/ JAO.ALMA\#2011.0.00396.S.}

\end{deluxetable}

\begin{deluxetable}{lllll}
\tabletypesize{\scriptsize}
\tablecolumns{5}
\tablecaption{ALMA Cycle-2 Mosaic Observations\label{tab:line}}
\tablewidth{\columnwidth}
\tablehead{
\colhead{Continuum \& Line}             & \colhead{Rest frequnecy} &
\colhead{$\Delta V$}    & \colhead{Beam Size (PA)}              & \colhead{noise level$^b$}    \\
\colhead{}                                  & \colhead{(GHz)} &
\colhead{(km s$^{-1}$)} & \colhead{arcsec$\times$arcsec ($^\circ$)} & \colhead{}  
}
\startdata
continuum  & 226$^a$ &  & 1.4 $\times$ 0.92 ($-$81.1)  & 0.024 (mJy beam$^{-1}$)   \\
$^{12}$CO ($J=2-1$)  & 230.538000 & 0.096 & 1.39 $\times$ 0.91 (89.0) & 0.02  (Jy beam$^{-1}$ km s$^{-1}$)  \\
$^{13}$CO ($J = 2-1$)  & 220.398684 & 0.096 & 1.45 $\times$ 0.94 (88.0) & 0.02  (Jy beam$^{-1}$ km s$^{-1}$)  \\
C$^{18}$O ($J = 2-1$)  & 219.560358 & 0.096 & 1.45 $\times$ 0.94 (88.0)& 0.015 (Jy beam$^{-1}$ km s$^{-1}$)  
\enddata
\tablenotetext{a}{Here are averaged frequencies of the center
frequencies of two basebands assigned for continuum observations
(218 GHz and 234 GHz). This corresponds to 1.3 mm in wavelength.}
\tablenotetext{b}{noise levels are measured for a 1 km s$^{-1}$ channel.}
\end{deluxetable}

\begin{deluxetable}{ccccccc}
\tabletypesize{\scriptsize}
\tablecaption{Measured Properties of SM1-A and Source-X\label{tab:properties}}
\tablewidth{0pt}
\tablehead{
\colhead{source} & \colhead{telescope} & \colhead{wavelength} & \colhead{Total Flux Density} & \colhead{Peak Flux Density} &\colhead{Beam Size (PA)} & \colhead{reference/comment} \\
\colhead{ } & \colhead{ } & \colhead{[$\mu m$]}　& \colhead{[mJy]}　& \colhead{[mJy beam$^{-1}$]} &\colhead{ [\arcsec (\arcdeg)]}  & 
}
\startdata
SM1-A & Chandra$^a$ & (2--10 keV) & 23 counts; 13 $\sigma$ & --  & 1.5   & this paper  \\
 & IRTF$/$SIRIUS & 2.2 & -- &  $<$ 0.011  & 0.5    &      \\
 & Spitzer$/$IRAC & 3.6 & -- & $<$0 .0044 & 1.66  &  $3 \sigma $ upper limit     \\
 & Spitzer$/$IRAC & 4.5 & -- & 0.031$\pm$0.01 & 1.72  &  marginal detection    \\
 & Spitzer$/$IRAC & 5.8 & -- & 0.16$\pm$0.05 & 1.88  &  marginal detection  \\
 & Spitzer$/$IRAC & 8.0 & -- & $<$ 0.31 & 1.98  &  $3 \sigma $ upper limit     \\
 & Spitzer$/$MIPS & 24 & -- & $<$ 2.9 & 6 &   $3 \sigma $ upper limit    \\
 & Herschel$/$PACS & 70 & -- & $<$ 1460 & 5.86 &  $3 \sigma $ upper limit     \\
 & ALMA/B7 & 810 & 366$\pm$4.6  & 264 $\pm$1.15 & 0.59 $\times$ 0.42 ($-$75) &      \\
 & ALMA/B7 & 835 & 327$\pm$1.4  & 245 $\pm$0.39 & 0.64 $\times$ 0.46 ($-$77) &      \\
 & SMA & 870 & 350$\pm$10$^b$  & 336 $\pm$10.5$^b$ & 2.8 $\times$ 0.9 (50.3) &      \\
 & ALMA/B6 & 1300 & 118.3$\pm$2.0$^c$  & 116.0 $\pm$0.11$^c$ & 1.4 $\times$ 0.92 ($-$81.1) &      \\
 & NMA & 2000 & --  & 84 $\pm$24 & 5.2 $\times$ 13.4 (--) &  3    \\
 & NMA & 3000 & --  & 24 $\pm$1.8 & 6.2 $\times$ 3.2 (--) &  3    \\
 & JVLA & 7300 & 0.608$\pm$0.065  & 0.443 $\pm$0.038 & 0.32 $\times$ 0.15 (11.4) &      \\
 & JVLA & 40000 & 0.125$\pm$0.056  & 0.081 $\pm$0.017 & 1.24  $\times$ 0.66 ($-$7.6) &  4    \\
 & JVLA & 66000 & 0.101$\pm$0.046  & 0.089 $\pm$0.022 & 2.0  $\times$ 1.0 ($-$4.8) &  4    \\
Source-X & Chandra$^a$ & (2--10 keV) & 35 counts; 19 $\sigma$ & --  & 1.5  &  this paper \\
 & ALMA/B6 & 1300 & 36.0$\pm$0.78  & 21.2 $\pm$0.29 & 0.6 $\times$ 0.34 ($-$89.27) &  219 GHz    \\
 & ALMA/B6 & 1300 & 37.9$\pm$ 1.1  & 35.8 $\pm$0.29 & 1.4 $\times$ 0.92 ($-$81.1) &  226 GHz    \\
  & ALMA/B3 & 3000 & 7.56   $\pm$ 0.86& 7.10$\pm$0.44 & 3.5 $\times$ 1.8 ($-$71) &  107 GHz, 5    \\
 & JVLA & 7300 & --  &  $<$ 0.114 & 0.32 $\times$ 0.15 (11.4) &   $3 \sigma $ upper limit   \\
 & JVLA & 40000 &  -  &  $<$ 0.051 & 1.24 $\times$ 0.66 ($-$7.6) & 4, $3 \sigma $ upper limit      \\
 & JVLA & 66000 & -   & $<$ 0.066 & 2.0 $\times$ 1.0 ($-$4.8) & 4,  $3 \sigma $ upper limit    
\enddata
\tablecomments{ Integrated and peak flux densities for images with ALMA and JVLA  were obtained with Gaussian fitting to the sources. }
\tablenotetext{a}{
Based on the X-ray observations, SM1-A and Source-X are also named as A-31 (J162627.8-242359) and A-29 (J162627.4-242418), 
respectively,  (Imanishi et al. 2003; Gagn\'e et al. 2004).}
\tablenotetext{b}{Peak flux density was obtained from the SMA image made using visibilities with u $>$ 20 $k\lambda$.}
\tablenotetext{c}{Integrated and peak flux densities for images at 226 GHz (made using 12m and 7m arrays) were also obtained with 
the Gaussian fitting task including ``sky subtraction" in CASA to the sources in order to remove contributions from extended structures.}
\tablerefs{
(1) Imanishi et al.  (2003), (2) Gagn\'e et al. (2004), (3) Kamazaki et al. (2001),
(4) Dzib et al. (2013), (5) Kirk et al. (2017)}
\end{deluxetable}

\clearpage

\begin{deluxetable}{ccccc}
\tablecaption{Physical Properties of SM1-A and Source-X\label{tab:properties2}}
\tablewidth{0pt}
\tablehead{
\colhead{property} & \colhead{unit} & \colhead{SM1-A} & \colhead{Source-X} & \colhead{First Core$^a$ }  }
\startdata
Mean gas density & cm$^{-3}$  & (2.2 -- 8.4) $\times$ 10$^{11}$ & 6.0 $\times$ 10$^{10}$ &   \\
Density &    g cm$^{-3}$ & (5.2 -- 33) $\times$ 10$^{-13}$ & 2.4 $\times$ 10$^{-13}$  & $>$ 10$^{-13}$     \\
Radius &  au & 21 & 27  & 4 -- 100    \\
Mass &  $M_\odot$ & 0.028 -- 0.14  & 0.018 -- 0.039  & $ \sim$ 0.1   \\
Temperature &  K & $ \sim$ 40 &  $ \sim$ 20 & $ \sim$ 100     \\
Dust Opacity index &  -- & 1.5 -- 2 & $>$ 1.4   & $ \sim$ 2     \\
Luminosity &  $L_\odot$ & 0.035 & ($ \sim$ 0.01) & 0.01-0.1     
\enddata
\tablenotetext{a}{The properties of first cores listed are from Larson (1969), Masunaga et al. (1998), and Saigo et al.  (2008)} 
\end{deluxetable}

\clearpage

\begin{deluxetable}{ccccc}
\tablecaption{Outflow Properties of SM1-A and Source-X\label{tab:outflow}}
\tablewidth{0pt}
\tablehead{
\colhead{property} & \colhead{unit} & \colhead{SM1-A} & \colhead{Source-X} & \colhead{BD model$^a$}  
}
\startdata
$^{12}$CO(2-1) Intensity & Jy km s$^{-1}$  &  7.34 $\pm$ 0.28& 1.14 $\pm$ 0.097  &   \\
$^{13}$CO(2-1) Intensity &  Jy km s$^{-1}$    & 2.7 $\pm$ 0.14 &  $<$ 0.1 &    \\
C$^{18}$O(2-1) Intensity &  Jy km s$^{-1}$    & 0.26 $\pm$ 0.02 & $<$ 0.15&    \\
$^{12}$CO(3-2) Intensity &  Jy km s$^{-1}$  & 26 $\pm$ 0.55  & --  &     \\
$^{12}$CO(2-1) $T_b$ peak & K  & 62 $\pm$ 1.6 & 11.5 $\pm$ 1.1   &    \\
$^{12}$CO(3-2) $T_b$ peak &   K & 104 $\pm$ 8.1 & -- &     \\
Outflow Velocity & km s$^{-1}$  & $\sim$ 3  & $\sim$ (5 -- 10)$^b$ &  $\sim$ 2 km s$^{-1}$   \\
Size &  au &  274 & 571  &  $\sim$ 350 au$^d$   \\
Dynamical Time &  yr & 434 & $\sim$ (285--628)$^b$ &  500$^d$  \\
Outflow Mass &  $M_\odot$ &  (1.3-3.9) $\times$ 10$^{-4}$ & (1.6--2.5) $\times$ 10$^{-5}$$^c$ & $\sim$ 2 $\times$ 10$^{-3}$    \\
Mass Loss Rate &  $M_\odot$ yr$^{-1}$ & (0.3-0.7) $\times$ 10$^{-6}$ &   (0.2 -- 0.7) $\times$ 10$^{-7}$ & $\sim$ 4 $\times$ 10$^{-6}$     
\enddata
\tablecomments{$^{13}$CO image around Source-X is rather affected by the strong emission in the vicinity of VLA 1623 and SM1 
especially at $V=$ 2km s$^{-1}$, and the upper limit to 
the $^{13}$CO intensity of the outflow component seen in $^{12}$CO is a tentative value obtained 
from the rms noise at $V_{\rm lsr} =$ 7 km s$^{-1}$.    
}
\tablenotetext{a}{Quantities listed here are taken from Machida et al. (2009) in which a proto-brown dwarf with $\sim$ 50 $M_{\rm jup}$ has an age of $<$ 500 yrs.}
\tablenotetext{b}{ Obtained for $i$ =  $\pm$ (15--30) deg. for both blue- and re-shifted components.}
\tablenotetext{c}{ Obtained using $^{12}$CO ($J=2-1$)  intensity assuming $T_{\rm ex}$ =30--60 K and the opacity is $\sim$ 10. }
\tablenotetext{d}{ We assumed the physcial quantities at the age of 500 yrs.}

\end{deluxetable}


\begin{thebibliography}{}
\bibitem[Alves et al.(2001)]{alves01}
Alves, J. F., Lada, C. J. \& Lada E. A. 2001, \nat, 409, 159
\bibitem[Alves de Oliveira et al.(2013)]{alves13}
Alves de Oliveira, C., \'Abrah\'am, P., Marton, G., et al. 2013, \aap, 559, A126
\bibitem[Anders \& Ebihara(1982)]{anders82}
Anders, E. \& Ebihara, M., 1982, , Geochimica et Cosmochimica Acta 46, 2363
\bibitem[Andr\'e et al. (1993)]{andre93} Andr\'e, F., Ward-Thompson, D.,  \& Barsony, M. 1993, \apj, 406, 122
\bibitem[Andr\'e et al. (2009)]{andre09} Andr\'e, F., Basu, S.,  \& Inutsuka, S.  2009, Structure Formation in Astrophysics, G. Chabrier, Cambridge: Cambridge Univ., 254
\bibitem[Balucinska-Church \& McGammon(1992)]{Balucinska92}
Balucinska-Church, M. \& McGammon, D. 1992, \apj, 400, 699
\bibitem[Basu \& Vorobyov (2012)]{basu12} Basu, S.,  \& Vorobyov, E.  2012, \apj, 750, 30
\bibitem[Bate et al.(2002)]{bate02} 
Bate, M., Bonnell, I.
    \& Bromm, V.  2002, \mnras, 322, L65
\bibitem[Bate(2010)]{bate10} Bate, M. 2010, \mnras, 404, 79
\bibitem[Bjerkeli et al.(2012)]{bjerkeli12}
Bjerkeli, P. et al. 2012, \aap, 549, A29
\bibitem[Bonnell et al.(2000)]{bonnell00}
Bonnell, I. A., Bate, M. R., Clarke, C. J., \& Pringle, J. E. 2000, \mnras, 323, 785
\bibitem[Di Francesco et al.(2004)]{difrancesco04} 
Di Francesco, J., Andr\' e, P. \& Myers, P. C., 2004, \apj, 617, 425
\bibitem[Chen et al.(2010)]{chen10}
Chen, X. et al., 2010, \apj, 715, 1344
\bibitem[Dzib et al.(2013)]{diz13} 
Dzib, K. et al. 2013, \apj, 755, 63
\bibitem[Enoch et al.(2010)]{enoch10}
Enoch, M. L. et al., 2010, \apj, 722, L33
\bibitem[Gagn\'e et al.(2004)]{gag04} Gagn\'e, M., Skinner, S., \& Daniel, K. 2004, \apj, 613, 392
\bibitem[Fazio et al. (2004)]{faz04} Fazio, G.G. et al. 2004, \apjs, 154, 10
\bibitem[Friesen et al. (2014)]{frie14} Friesen, R., Di Francesco, J., Bourke, T. L. et al. 2014, \apj, 797, 27
\bibitem[Hara et al. (2013)] {hara2013} Hara, C. et al. 2013,
\apj, 771, 128
\bibitem[Haris et al. (2018)] {harris18} Haris, R. et al. 2018, \apj, 861, 91
\bibitem[Hayashi(1966)]{hayashi66}
Hayashi, C. 1966, \araa, 4, 171
\bibitem[Hildebrand(1983)]{hilde83}
Hildebrand, R. H. 1983, Q. J. R. Astron. Soc. 24, 267
\bibitem[Ho et al.(2004)]{ho04}
Ho, P.l T. P., Moran, J. M., Lo, K. Y., 2004, \apj, 616, 1
%
\bibitem[Imanishi et al. (2001)]{ima01}Imanishi, K. et al. 2001, \apj, 557, 747
\bibitem[Imanishi et al. (2003)]{ima03} Imanishi, K. et al. 2003, \pasj, 55, 653
\bibitem[Inutsuka et al.(2010)]{inu10}
Inutsuka, S., Machida, M., \& Matsumoto, T. 2010, \apj, 718, L58
\bibitem[Kamazaki et al. (2001)]{kama01} Kamazaki, T. et al.,   2001, \apj, 548, 278
\bibitem[Kamazaki et al. (2003)]{kama03} Kamazaki, T. et al.,   2003, \apj, 584, 375
\bibitem[Kamazaki et al. (2018)]{kama18} Kamazaki, T. et al.,   2018, submitted to \apj
\bibitem[Kirk et al.(2017)]{kirk17}
Kirk, H., Dunham, M. M., Di Francesco, J., et al. 2017, \apj, 838, 114
\bibitem[Kunde \& Hog(1998)]{kunde98}
Knude, J., \& Hog, E. 1998, \aap, 338, 897
\bibitem[Larson (1969)]{lar69} Larson, B. R. 1969, \mnras, 145, 271
\bibitem[Loinard et al.(2008)]{loinard08}
Loinard, L., Torres, R. M., Mioduszewski, A. J., \& Rodriguez, L. F. 2008, \apj, 675, L29
\bibitem[Lombardi et al.(2008)]{lombardi08}
Lombardi, M., Lada, C. J., \& Alves, J. 2008, \aap, 480, 785
\bibitem[Luhman et al., (2007)]{luh09} 
Luhman, K.L. et al. 2007, Protostar and Planets V, B. Reipurth, D. Jewitt, and K. Keil, Arizona: The university of Arizona Cambridge Press, 443
\bibitem[Machida et al.(2008)]{machi08} Machida, M., Inutsuka S., 
    \& Matsumoto, T.  2008, \apj, 676, 1088
\bibitem[Machida et al.(2009)]{machi09} Machida, M., Inutsuka S., 
    \& Matsumoto, T.  2009, \apj, 699, L157
    \bibitem[Matsumoto \& Hanawa(2003)]{matsu03} 
    Matsumoto, T.  \& Hanawa, T. 2003, \apj, 595, 913
 \bibitem[Masunaga et al.(1998)]{masu98} Masunaga, H, Miyama, S., \&  Inutsuka S. 
     1998, \apj, 495, 346
 \bibitem[Morrison \& McCammon(1983)]{morrison83}
Morrison, R. \& McCammon, D. 1983, \apj, 270, 119
 \bibitem[McKee \& Ostriker(2007)]{mckee07} 
McKee, C. F. \& Ostriker, E. C., 2007, araa, 45, 565
 \bibitem[Murillo et al.(20133)]{murillo13} Murillo, N. M., Lai, S.-P., Bruderer, S. et al. 
     2013, \aap, 560, 103
\bibitem[Nakamura  et al.(2012)]{naka12} Nakamura, F., Takakuwa, S.
    \& Kawabe, R.  2012, \apjl, 758, L25 
    \bibitem[Ortiz-Leon et al. (2017)]{ortiz17}
Ortiz-Leon, G. N., Loinard L., Kounkel, M. A., et al. 2017, \apj, 834, 141
\bibitem[Pineda et al.(2011)]{pineda11}
Pineda, et al. 2011, \apj, 743, 2011
\bibitem[Reipurth \& Clarke (2001)]{reipurth01} Reipurth, B., \& Clarke C., 2001, \aj, 122, 432
\bibitem[Rieke et al. (2004)]{rie04} Rieke, G.H. et al. 2004, \apjs, 154, 25
\bibitem[Sault et al.(1995)]{sau95} Sault, R. J., Teuben, P. J., \& Wright, M. C. H.\ 1995,
Astronomical Data Analysis Software and Systems IV, ASP
Conference Series, Vol. 77, R. A. Shaw, H. E. Payne, \& J. J. E. Hayes, eds., p.433
\bibitem[Saigo et al.(2008)]{saigo08} Saigo, K., Tomisaka, K.
    \& Matsumoto, T.  2008, \apj, 674, 997
\bibitem[Saigo \& Tomisaka (2011)]{saigo11} Saigo, K. \& Tomisaka, K.
     2009, \apj, 728, 78 
\bibitem[Scoville et al.(1993)]{sco93} Scoville, N. Z., Carlstrom, J. E., Chandler, C. J., et al.\ 1993, \pasp, 105, 1482
\bibitem[Shu et al.(1987)]{shu87}
      Shu, F. H., Adamas, F. C. \& Lizano, S., 1987, \araa, 25, 23
      \bibitem[Smith et al.(2001)]{smith01}
Smith, R., K., Brickhouse, N. S., Liedahl, D. A.  et al.,  2001, \apj, 556, L91
 \bibitem[Stamatellos, \& Whitworth (2009)]{stama09}
 Stamatellos, D., \& Whitworth, A.
     2009, \mnras, 392, 413  
\bibitem[Tomida et al. (2010)]{tomi10} Tomida, K. et al. 2010, \apjl, 725, L239 
\bibitem[Tomida et al. (2013)]{tomi13} Tomida, K. et al. 2013, \apj, 763, 6
\bibitem[Tomisaka(2002)]{tomi02}
Tomisaka, K., 2002, \apj, 575, 306
\bibitem[Ward-Thompson et al. (1989)]{ward-thompson89}
Ward-Thompson, D., Robson, E. I., Whittet, D. C. B. et al. 1989, \mnras, 241, 119 
\end{thebibliography}
\end{document}